%% file: DustEB.tex
\documentclass[twocolumn,traditabstract,longauth,a4]{aa}
\usepackage{natbib}
\usepackage{url}
\usepackage{graphicx}
\usepackage{subfigure}
\usepackage{amsmath}
\usepackage{txfonts}
\usepackage{color}
\usepackage{listings} 
\usepackage{courier}
\usepackage{multirow}
\usepackage{alltt}
\usepackage{float}
\usepackage{morefloats}
\usepackage{ifthen}
\usepackage{longtable}
\bibpunct{(}{)}{;}{a}{}{,} 
\usepackage[colorlinks=true,linkcolor=red,citecolor=blue, breaklinks=true]{hyperref}
\usepackage[switch,displaymath, mathlines,]{lineno}
\usepackage{tikz}

\input Planck.tex

\DeclareRobustCommand{\rchi}{{\mathpalette\irchi\relax}}
\newcommand{\irchi}[2]{\raisebox{\depth}{$#1\chi$}} 

\newcommand{\healpix}{\ensuremath{\tt HEALPix}}
\newcommand{\disperse}{\ensuremath{\tt DisPerSE}}
\newcommand{\getfilaments}{\ensuremath{\tt getfilaments}}
\newcommand{\smaff}{\ensuremath{\tt SMAFF}}

\newcommand{\isynfast}{\ensuremath{\tt isynfast}}
\newcommand{\ianafast}{\ensuremath{\tt ianafast}}
\newcommand{\gnomview}{\ensuremath{\tt gnomview}}
\newcommand{\neighbourring}{\ensuremath{\tt neighbours\_ring}}
\newcommand{\planck}{{\it Planck\/}}
\newcommand{\EB}{{$E$-$B$}}
\newcommand{\herschel}{{\it Herschel\/}}
\newcommand{\wmap}{{WMAP\/}}
\newcommand{\iras}{{IRAS\/}}
\newcommand{\hi}{{\sc Hi}}

\newcommand{\mukcmb}{\ensuremath{\mu{\rm K}_{\rm CMB}}}
\newcommand{\musqkcmb}{\ensuremath{\mu{\rm K}_{\rm CMB}^2}}
\newcommand{\Nside}{\ensuremath{N_{\rm side}}}
\newcommand{\Dmodel}{\ensuremath{D_{\rm 353}}}
\newcommand{\fDmodel}{\ensuremath{D_{\rm 353}^{\rm b}}}
\newcommand{\fE}{\ensuremath{E_{\rm 353}^{\rm b}}}
\newcommand{\fB}{\ensuremath{B_{\rm 353}^{\rm b}}}
\newcommand{\StokesI}{{\ifmmode I\else $I$\fi}}     
\newcommand{\StokesQ}{{\ifmmode Q\else $Q$\fi}}     
\newcommand{\StokesU}{{\ifmmode U\else $U$\fi}}     
\newcommand{\Q}{\ensuremath{Q_{\rm 353}}}    
\newcommand{\U}{\ensuremath{U_{\rm 353}}}    
\newcommand{\bQ}{\ensuremath{Q^{\prime}_{\rm 353}}}    
\newcommand{\bU}{\ensuremath{U^{\prime}_{\rm 353}}}    
\newcommand{\fQ}{\ensuremath{Q^{\prime\rm b}_{\rm 353}}}    
\newcommand{\fU}{\ensuremath{U^{\prime\rm b}_{\rm 353}}}    
\newcommand{\aE}{\ensuremath{\bar{E}_{\rm 353}}}    
\newcommand{\aB}{\ensuremath{\bar{B}_{\rm 353}}}    
\newcommand{\bFD}{\ensuremath{D^{\rm F}_{\rm 353}}} 
\newcommand{\bBD}{\ensuremath{D^{\rm B}_{\rm 353}}} 
\newcommand{\bFQ}{\ensuremath{Q^{\prime \rm F}_{\rm 353}}}    
\newcommand{\bFU}{\ensuremath{U^{\prime \rm F}_{\rm 353}}}    
\newcommand{\bBQ}{\ensuremath{Q^{\prime \rm B}_{\rm 353}}}    
\newcommand{\bBU}{\ensuremath{U^{\prime \rm B}_{\rm 353}}}    
\newcommand{\abFQ}{\ensuremath{\bar{Q}^{\prime \rm F}_{\rm 353}}}    
\newcommand{\abFU}{\ensuremath{\bar{U}^{\prime \rm F}_{\rm 353}}}    
\newcommand{\E}{\ensuremath{E_{\rm 353}}}    
\newcommand{\B}{\ensuremath{B_{\rm 353}}}    
\newcommand{\EE}{{\ifmmode EE\else $EE$\fi}}    
\newcommand{\BB}{{\ifmmode BB\else $BB$\fi}}    
\newcommand{\TE}{{\ifmmode TE\else $TE$\fi}}    
\newcommand{\TB}{{\ifmmode TB\else $TB$\fi}}    
\newcommand{\mlambda}{\ensuremath{\lambda_{-}}}
\newcommand{\mtheta}{\ensuremath{\theta_{-}}}
\newcommand{\hessangle}{\ensuremath{\bar{\theta}_{-}}} 
\newcommand{\angloc}{\ensuremath{\bar{\rchi}}} 
\newcommand{\anggmf}{\ensuremath{\bar{\rchi}_{\rm m}}} 
\newcommand{\bpos}{\ensuremath{\vec{B}_\mathrm{POS}}}
\newcommand{\bgmf}{\ensuremath{\vec{B}_\mathrm{GMF}}}
\newcommand{\bmean}{\ensuremath{\vec{B}_{\mathrm{m,POS}}}}

\begin{document}

\input PIP_118_Boulanger_authors_and_institutes
\title{{\Planck} intermediate results. XXXVIII. $E$- and $B$-modes of dust polarization from the magnetized filamentary structure of the  interstellar medium}

\abstract{The quest for a $B$-mode imprint  from primordial gravity waves on the polarization of the cosmic microwave background (CMB) requires the characterization of foreground polarization from 
Galactic dust. We present a statistical study of the filamentary structure of the $353\,$GHz \Planck\  Stokes maps at high Galactic latitude, relevant to the study of dust emission as a polarized foreground to the 
CMB. We filter the intensity and polarization maps to isolate filaments in the range of angular scales where the power asymmetry between $E$-modes and $B$-modes is observed. Using the Smoothed Hessian Major Axis Filament Finder (\smaff), we identify 
259 filaments at high Galactic latitude, with lengths larger or equal to $2$\deg\ (corresponding to 3.5\,pc in length for a typical distance of 100\,pc). These filaments show a preferred orientation parallel to the magnetic field 
projected onto the plane of the sky, derived from their polarization angles. We present mean maps of the filaments in Stokes \StokesI, \StokesQ,  \StokesU,  $E$, and $B$, computed by stacking individual images 
rotated to align the orientations of the filaments. Combining the stacked images and the histogram of relative orientations, we estimate the mean polarization fraction of the filaments to be 11\,\%. Furthermore, we show that 
the correlation between the filaments and the magnetic field orientations may account for the $E$ and $B$ asymmetry and the $C_{\ell}^{TE}/C_{\ell}^{EE}$ ratio, reported in the power spectra analysis of the \Planck\ $353$\,GHz polarization 
maps. Future models of the dust foreground for CMB polarization studies will need to take into account the observed correlation between the dust polarization and the 
structure of interstellar matter.}

\titlerunning{$E$- and $B$-modes of dust polarization} 
\authorrunning{\Planck\ Collaboration} 
\maketitle


\clearpage

\section{Introduction} \label{sec:intro}

Recently, \Planck\footnote{\Planck\ (\url{http://www.esa.int/Planck}) is a project of the European Space Agency (ESA) with instruments provided by two scientific consortia funded by ESA member states and led 
by Principal Investigators from France and Italy, telescope reflectors provided through a collaboration between ESA and a scientific consortium led and funded by Denmark, and additional contributions from 
NASA (USA).}  has reported an asymmetry in power between the dust $E$- and $B$-modes in its 353\,GHz observations  \citep{planck2014-XXX, planck2014-a12}. This power asymmetry has been observed outside
masks covering 20 to 70\,\% of the sky, excluding the Galactic plane. The ratio of the dust $B$- to $E$-mode power amplitudes is about a half over the multipole range $40 < \ell < 600$ \citep{planck2014-XXX}. The 
source of this power asymmetry in the dust polarization data is currently unknown. Models of the Galactic magnetic field (GMF) used in the Planck sky model (PSM, \citealt{PSM:2013}) and the FGPol model 
\citep{ODea:2012} produce an equal amount of power in $E$- and $B$-modes outside the regions covered by the sky masks. These models, which were used to estimate the dust polarization foreground 
\citep{BICEP2:2014,pb2015}, include an analytical model of the large-scale GMF (\bgmf) and a statistical description of the turbulent component of the magnetic field. 

The \planck\ maps of thermal dust emission display filaments distributed over the whole sky \citep{planck2013-p06b}. The filamentary structure of the diffuse interstellar matter is also a striking feature of dust 
observations at higher angular resolution, performed by  \herschel, and of spectroscopic \hi\ observations  \citep[e.g.][]{MAMD:2010,Andre:2014,Clark:2014}. The analysis of \planck\ dust polarization data in the 
diffuse interstellar medium (ISM), at low and intermediate Galactic latitudes, indicates that the structures of interstellar matter tend to be aligned with the plane of the sky (POS) projection of the magnetic field (\bpos, \citealt{planck2014-XXXII}). 
This preferential relative orientation is also observed in simulations of magneto-hydrodynamic (MHD) turbulence of the diffuse ISM \citep{Hennebelle:2013, Soler:2013}. Such a coupling between the structure of 
interstellar matter and \bpos\ is not included in the PSM or FGPol models of the dust polarization sky \citep{PSM:2013, ODea:2012}.

The goal of this paper is to test whether the correlation between the filamentary structures of the intensity map and \bpos\ in the diffuse ISM accounts for the observed \EB\ asymmetry. \cite{Zaldarriaga:2001} 
describes the $E$- and $B$-modes decomposition of simple patterns of polarized emission, including filaments with a homogeneous polarization degree and orientation. The presence of $E$-modes is related to 
invariance by parity of the polarization pattern. There is $E$-only power if \bpos\ is either parallel or perpendicular to the  filaments. If \bpos\ is oriented at $+ 45$\deg\ or $- 45$\deg\ with respect to the filaments, 
there is $B$-only power. 

In this paper, we filter the \planck\ intensity and polarization maps to isolate filaments in the range of angular scales where the \EB\ asymmetry is observed. We identify coherent elongated filaments within regions of low column 
density at high Galactic latitude using a filament-finding algorithm. We evaluate the mean polarization angle in each of these filaments and compare it to the mean orientation of each filament. In doing so, we extend the 
analysis presented in \citet{planck2014-XXXII} to  the relevant region of the sky for CMB polarization observations at high Galactic latitude. In order to achieve a high signal-to-noise ratio and enhance the contrast with 
respect to the local background dust emission, we stack the Stokes  \StokesI, \StokesQ,  \StokesU, and also $E$ and $B$ maps, for the filaments we select in the \Planck\ dust intensity map. We use the stacked images to 
quantify the power asymmetry in $E$- and $B$-modes associated with the filaments.

This paper is organized as follows. In Sect.~\ref{sec:data}, we introduce the \planck\ $353$\,GHz data used in this study. The filament-finding algorithm is presented in Sect.~\ref{sec:algorithm}. 
Section~\ref{sec:hro} presents the study of relative orientation between the filaments, \bpos, and the POS component of the large-scale GMF (\bmean) at high Galactic latitude. In Sect.~\ref{sec:pol_frac}, we 
present the stacking of both intensity and polarization maps and derive the mean polarization fraction of the filaments. In Sect.~\ref{sec:eb_asymmetry}, we discuss the relation between the relative orientation of the filaments 
and \bpos\ and the \EB\ asymmetry. Section~\ref{sec:gal_astrophysics} presents our results in the context of earlier studies and its relation to Galactic astrophysics. Finally, we present our conclusions in Sect.~\ref{sec:conclusion}. This paper has four appendices. Appendix~\ref{sec:hessian} details the Hessian analysis implemented to identify the 
filaments in the dust intensity map. The application of the filament-finding algorithm to a simulated  Gaussian dust sky is detailed in Appendix~\ref{sec:gauss_simul}. In Appendix~\ref{sec:vary_threshold}, we study the 
impact of our selection of the filaments on the main results of the paper. The computation of all the angle uncertainties that we use in our analysis is presented in Appendix~\ref{sec:angle_errors}.

\section{\Planck\ data} \label{sec:data}

The \planck\ satellite has observed the sky polarization in seven frequency bands from $30$ to $353$\,GHz \citep{planck2014-a01}. In this paper, we only use the 2015 (``DX11d") data from the High Frequency Instrument 
(HFI, \citealt{lamarre2010}) at 353\,GHz, since they are best suited to study the structure of dust polarization \citep{planck2014-XIX,planck2014-XX,planck2014-XXI,planck2014-XXII}. The data processing, 
map-making, and calibration of the HFI data are described in \citet{planck2014-a08} and \citet{planck2014-a09}. In our analysis, we ignore the dust and CO spectral mismatch leakage from intensity to polarization 
\citep{planck2014-a09}. \citet{planck2014-XXX} has shown that the amplitude of the dust spectral mismatch leakage at high latitude ($f_{\rm sky}=0.5$) is small compared to the total polarization signal in $E$ and $B$ 
modes. No CO emission is detected at 353\,GHz away from the Galactic plane and the brightest molecular clouds \citep{planck2013-p03a} and so we do not consider it in our analysis.

To quantify the statistical noise and systematic effects on the results presented in this paper, we use the two HalfMission (HM), two DetSet (DS), and two HalfRing (HR) \planck\ $353$\,GHz polarization maps \citep{planck2014-a08}. The 
two HM maps are made from the two halves of the full-mission \planck\ data, whereas the two HR  maps are produced by splitting each ring (also called stable pointing period) into two equal duration 
parts. The two DS maps are constructed using two subsets of polarization-sensitive bolometers at a given frequency. The noise is uncorrelated between the two HM, HR, and DS maps. We only use them to compute the error bars on the relevant quantities that we measure in this paper. 

The total polarization intensity ($P_{353}$) and the polarization angle ($\psi$) are derived from the full-mission Stokes \Q\ and \U\ maps at $353$\,GHz using the relations
\begin{align}
P_{353} &=  \sqrt{\Q^2 + \U^2}\ , \label{eq:2.1} \\
\psi & = 0.5 \times \text{atan2} (- \U, \Q) \ , \label{eq:2.2} 
\end{align} 
where the two-argument function $\text{atan2} (-\U, \Q)$ is used to compute $\text{atan} (-\U/\Q)$ avoiding the $\pi$ ambiguity. To recover the correct full range of polarization angles ( [$-\pi/2 , \pi/2$] as used for $\psi$ here), attention must be paid to the signs of both \U\ and \Q, not just their ratio. 
We use the IAU convention for $\psi$, which is measured from the Galactic North (GN) and 
positive to the East. The minus sign in Eq.~\eqref{eq:2.2} converts the convention provided in the \planck\ data to that of the IAU \citep[see][]{planck2014-XIX}. The orientation angle ($\rchi$) of \bpos\ is defined within the 
$\pi$ ambiguity by adding $\pi/2$ to the polarization angle
\begin{equation}
\rchi = \psi + \frac{\pi}{2} \ .  \label{eq:2.3} 
\end{equation}

For the dust intensity at 353\,GHz, we use the model map \Dmodel, computed from a modified blackbody fit to the \planck\ data at $\nu\,\ge\,353$\,GHz and  the \iras\ 100\,\micron\ map \citep{planck2013-p06b}. This map 
has lower noise than the $353$\,GHz Stokes \StokesI\ map and is corrected for zodiacal light emission, CMB anisotropies, and the cosmic infrared background monopole. We neglect the contribution of the CMB 
polarization at $353$\,GHz for this study. 

The full-mission \planck\ Stokes \Q\ and \U\ maps are provided in \healpix\footnote{\url{http://healpix.jpl.nasa.gov}} format \citep{Gorski:2005} at 4\parcm8 resolution and \Dmodel\ at 5\arcm. To increase the signal-to-noise 
ratio, we smooth the three maps to a common resolution of 15\arcm, taking into account the effective beam response of each map, and reduce to a \healpix\ resolution of $\Nside=512$. For the polarization data, we 
decompose the Stokes  \Q\ and \U\  maps into  \E\ and \B\ $a_{\ell m}$s ($E_{\ell m}$ and $B_{\ell m}$) using the ``\ianafast" routine of \healpix, apply the Gaussian smoothing in harmonic space (after deconvolving the effective azimuthally symmetric 
beam response of each map), and transform the smoothed  \E\ and \B\ $a_{\ell m}$s back to \Q\ and \U\  maps using the ``\isynfast" routine at $\Nside=512$. We also transform the \E\ and \B\ $a_{\ell m}$s to \E\ and \B\ maps at $\Nside=512$ using the relations
\begin{equation}
\E(\hat{\vec{\rm n}}) = \sum E_{\ell m} Y_{\ell m} (\hat{\vec{\rm n}}) , \ \ \ \B(\hat{\vec{\rm n}}) = \sum B_{\ell m} Y_{\ell m} (\hat{\vec{\rm n}}) \ .
\end{equation}
All the maps that we use are in thermodynamic units (\mukcmb).

In this paper, we work with the bandpass-filtered dust intensity map, \fDmodel, to identify and isolate filaments over the filtering scale using a filament-finding algorithm.
By filtering out large-scale and small-scale modes, we enhance the contrast of the filaments with respect to the 
diffuse background and reduce the instrumental noise, which is critical for accurately measuring the polarization orientations of the filaments within regions of low column density at high Galactic latitude.

\begin{figure}[t]
\includegraphics[width=8.8cm]{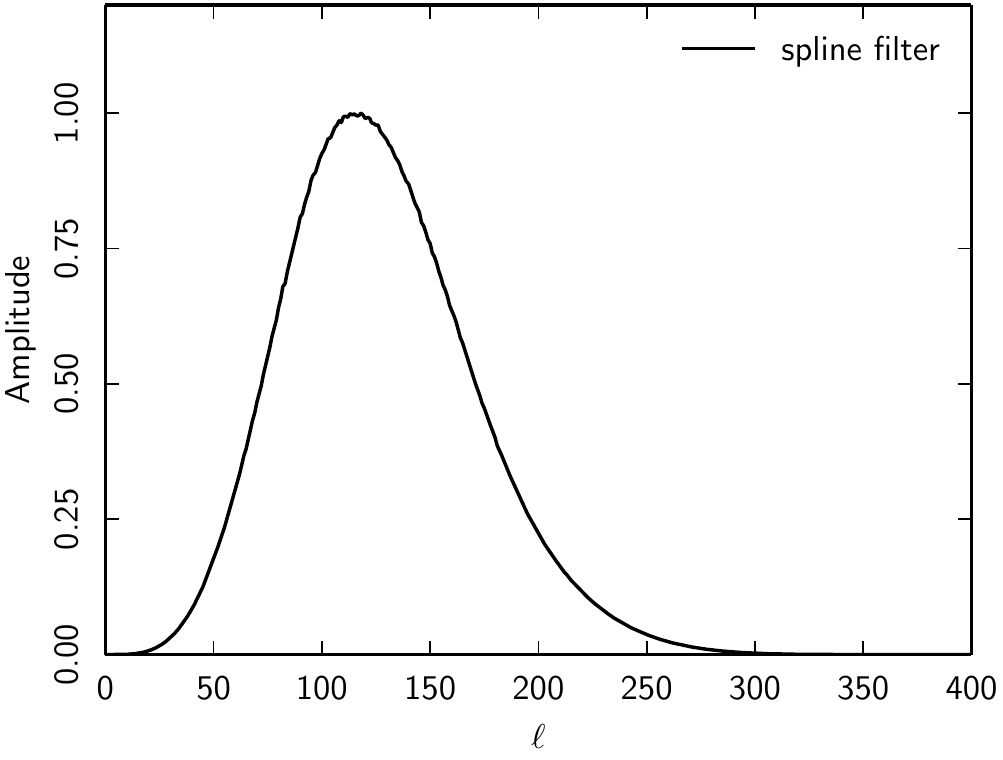}
\caption{The representative bandpass filter, retaining only the scales between $\ell=30$ and 300.}
\label{fig:2.1}
\end{figure}

For filtering, we apply the three-dimensional spline wavelet decomposition based on the undecimated wavelet transform, as described by \citet{Starck:2006}. We use the publicly available package Interactive Sparse 
Astronomical Data Analysis Packages (ISAP\footnote{\url{http://www.cosmostat.org/isap.html}}) to compute the \fDmodel\ map at $\Nside=512$ resolution. The spline wavelet used in this analysis provides less 
oscillation in position space compared to Meyers or needlet ones \citep{Lanusse:2012}. The filtering is done in pixel space; the corresponding bandpass filter in harmonic space varies a little over the 
sky. Figure~\ref{fig:2.1} presents the typical shape of this bandpass filter, which selects the scales between $\ell=30$ and 300. The filtering scale is chosen in such a way that it highlights all the bright filaments present in the \planck\ \Dmodel\ map. We also compute the bandpass-filtered polarization maps, $Q_{\rm 353}^{\rm b}$, $U_{\rm 353}^{\rm b}$, \fE, and \fB.

\begin{figure*}
\begin{tabular}{c}
\includegraphics[width=18cm]{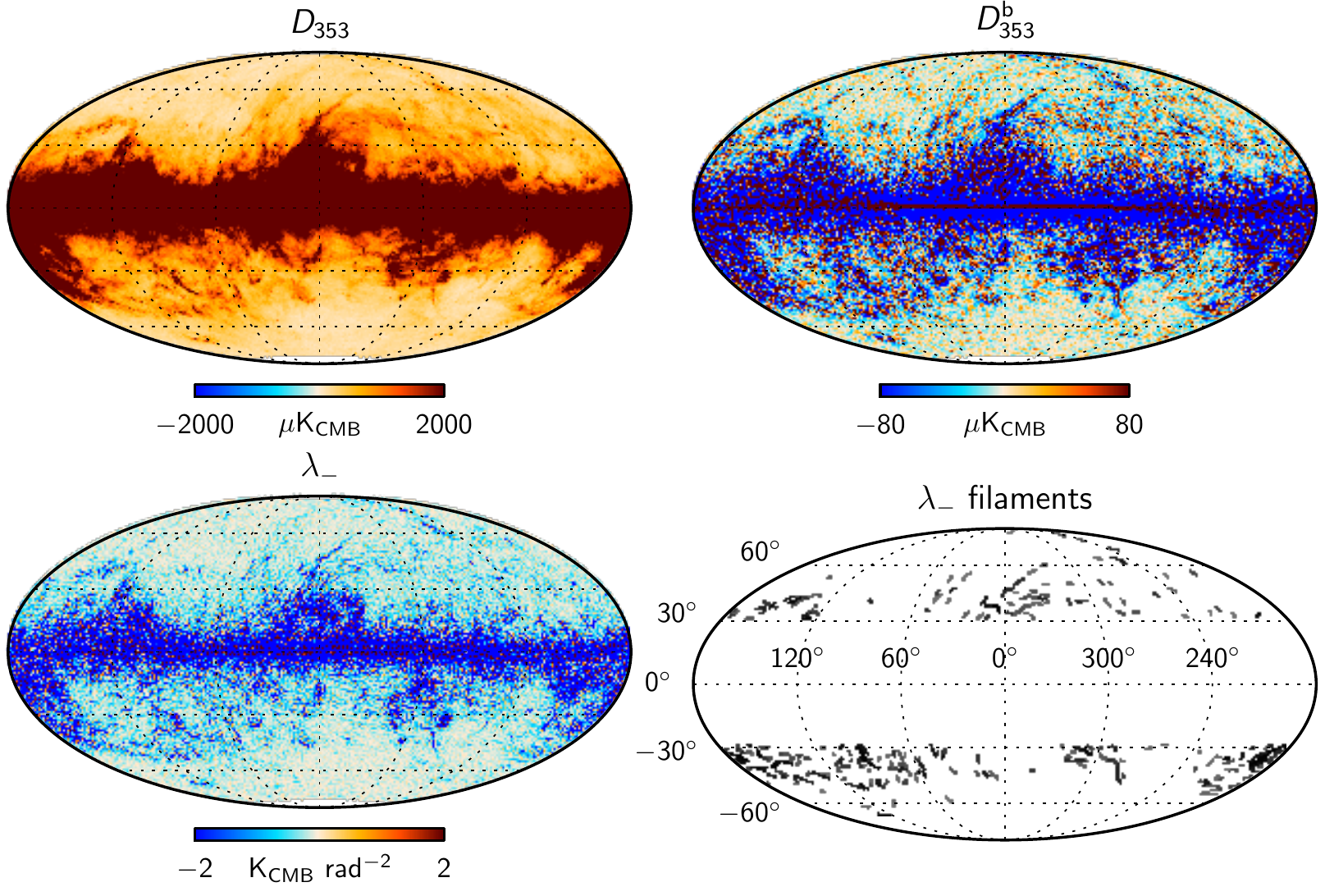}
\end{tabular}
\caption{Data processing steps implemented to identify filaments from the \planck\ data. We start with the \planck\ \Dmodel\ map (upper left panel) smoothed at 15\arcm\ resolution. The bandpass-filtered \fDmodel\ 
map (upper right panel) is produced using the spline wavelet decomposition, retaining only the scales between $\ell=30$ and 300. The lower eigenvalue map of the Hessian matrix, \mlambda, is shown in the 
lower left panel. Structures identified in the high-latitude sky  \mlambda\ map are shown in the lower right panel. The superimposed graticule is plotted in each image and labelled only on the 
lower right panel. It shows lines of constant longitude separated by 60\deg\ and lines of constant latitude separated by 30\deg. The same graticule is used in all plots of the paper. }
\label{fig:3.1}
\end{figure*}

\section{Filament-finding algorithm}\label{sec:algorithm}

\subsection{Methodology}\label{sec:3.1}

Identification of filaments as coherent structures is a crucial part of this analysis. Previous studies used algorithms such as \disperse\ \citep{Sousbie:2011,Arzoumanian:2011}, \getfilaments\  
\citep{Menshchikov:2013}, and the rolling Hough transform \citep{Clark:2014}. \citet{Hennebelle:2013} and \citet{Soler:2013} used the inertia matrix and the gradient of the density and column density 
fields to identify filaments in numerical simulations of MHD turbulence.

In this paper, we employ the Smoothed Hessian Major Axis Filament Finder (\smaff, \citealt{Bond:2010a}) algorithm, which has been used to identify filaments in the three-dimensional  galaxy distribution 
\citep{Bond:2010b}. \smaff\ is primarily based on the Hessian analysis. The Hessian analysis has also been used to analyse the \planck\ dust total intensity map in \cite{planck2014-XXXII},  \herschel\ images of 
the L1641 cloud in Orion A \citep{Polychroni:2013}, and large-scale structure in simulations of the cosmic web \citep{Colombi:2000,Forero-Romero:2009}. \citet{planck2014-XXXII} has reported good agreement  between the filament orientations derived from the Hessian and inertia matrix algorithms.

\subsection{Implementation}\label{sec:3.2}

In this study, we apply the two-dimensional  version of \smaff\ to the \fDmodel\ map, which is shown in the upper right panel of Fig.~\ref{fig:3.1}. From the Hessian matrix, we compute an all-sky map of the lower eigenvalue 
\mlambda\ and the orientation angle \mtheta\ of the perpendicular to the corresponding eigenvector, measured with respect to the GN. The details of the Hessian analysis are provided in Appendix~\ref{sec:hessian}. The map of \mlambda\ is presented in the lower left
panel of Fig.~\ref{fig:3.1}. For the subsequent analysis, we consider only the high-latitude sky, defined as $|b| > 30\deg$, with the Large Magellanic Cloud and Small Magellanic Cloud regions masked out. 

\begin{figure}[ht!]
\begin{tabular}{c}
\includegraphics[width=8.6cm]{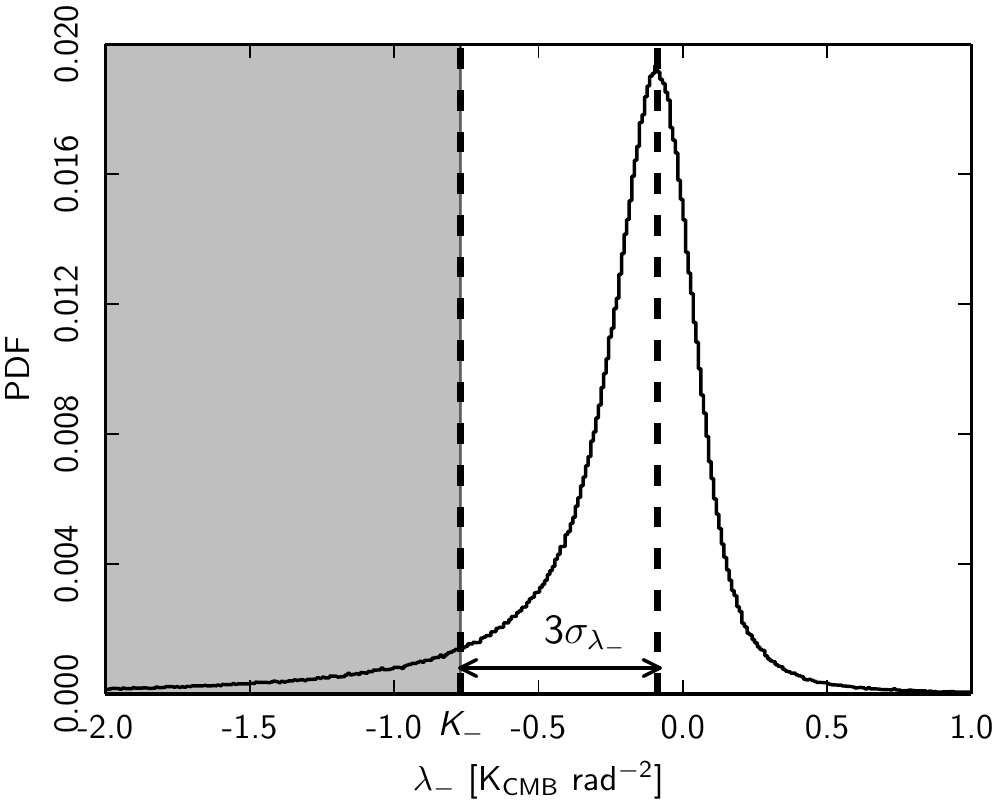}
\end{tabular}
\caption{Distribution of the eigenvalues \mlambda\ over the unmasked pixels in the high-latitude sky. The grey region represents the pixels that were used in the \smaff\ algorithm to find strong filaments.}
\label{fig:3.2}
\end{figure}

The map of  \mlambda\ highlights filaments in the \fDmodel\ map with an orientation angle \mtheta, which we refer to as the Hessian angle hereafter.  The distribution of \mlambda\ over the unmasked pixels is shown in 
Fig.~\ref{fig:3.2}. This distribution of \mlambda\ is non-Gaussian with an extended tail. We use the median absolute deviation (MAD, \citealt{Hampel:1974,Komm:1999}) to measure the width, $\sigma_{\lambda_{-}}$, 
of the distribution, as given by
\begin{equation}
\sigma_{\lambda_{-}} = 1.4826 \times \text{median} ({|\mlambda - m_{\mlambda}|)} \ , \label{eq:3.1}
\end{equation}
where $m_{\mlambda}$ is the median of the \mlambda\ distribution. We select filaments using an upper threshold ($K_{-}$) on \mlambda\ given by
\begin{equation} 
K_{-}= m_{\mlambda}- 3\,\sigma_{\lambda_{-}} \ . 
\end{equation}
Hereafter, we refer to the filaments satisfying $\mlambda< K_{-}$ as ``strong". This threshold $K_{-}$ separates the strong filaments from the weak ones, as detailed in Appendices~\ref{sec:gauss_simul} 
and~\ref{sec:vary_threshold}. By construction, the threshold $K_{-}$ rejects pixels where \mlambda\ is positive, since those pixels do not correspond to local maxima. 

We seek coherent elongated structures in the map. In \smaff, this is achieved by placing an upper limit $C$ on the difference between Hessian angles within a given structure. For our purpose, we set the value of $C=15\deg$ to identify relatively straight filaments.

We start with the pixel having the most negative \mlambda\ and denote the corresponding Hessian orientation angle by $\mtheta^{\rm s}$. We identify its neighbouring pixels using the ``\neighbourring'' routine 
of \healpix\  and look for pixels with $\mlambda  < K_{-}$ and orientation angle such that $|\mtheta-\mtheta^{\rm s}| \le C$.  If both conditions are satisfied, we count that neighbouring pixel as a part of the 
filament and move on to that neighbouring pixel. The neighbouring pixel becomes the new reference point and we search for its neighbours that satisfy both conditions. In our algorithm, $\mtheta^{\rm s}$ is 
fixed by the starting pixel, which has  the most negative \mlambda. We continue this friend-of-friend  algorithm to connect pixels until one of the conditions is no longer satisfied. We limit our selection to filaments 
with a length ($L$, defined as the maximum angular distance between pixels within a given structure) larger than or equal to the threshold length $L_{0}$, which we choose to be 2\deg. This process yields a set 
of 259 elongated filaments, as shown in the lower left panel of Fig.~\ref{fig:3.1}. Hereafter, we refer to this set as our filament sample. Selected sky pixels  represent 2.2\,\% of the high-latitude sky 
considered in our analysis.  There is no overlap between the filaments in our sample. 

\begin{figure}[ht!]
\includegraphics[width=8.8cm]{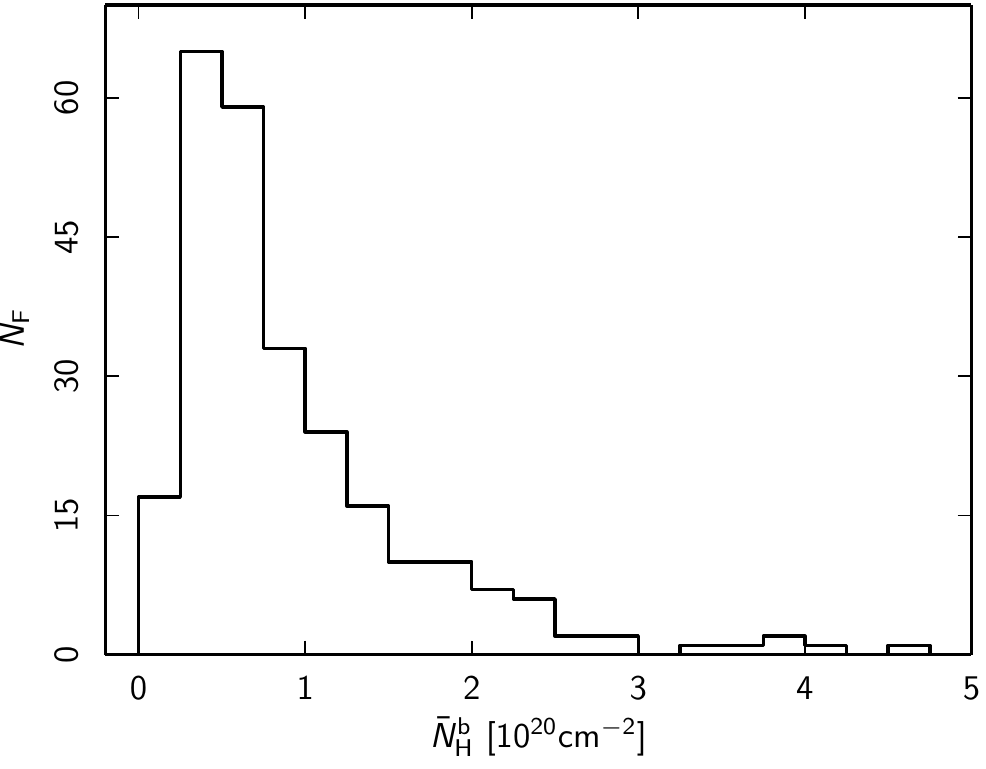}
\caption{Histogram of the mean column density of the filament sample. The column density is computed from the \fDmodel\ map using the conversion factor derived in \citet{planck2013-XVII}.}
\label{fig:3.3}
\end{figure}

The column density is computed from the \fDmodel\ map using the conversion factor, $\rm 0.039 \,MJy\,sr^{-1}$ per $10^{20}\,$H\,cm$^{-2}$; this was derived in \citet{planck2013-XVII} by correlating the \Planck\ 
$353\,$GHz dust total emission map with an \hi\ column density map over the southern Galactic polar cap. We average the column density along each filament and assign one mean column 
density, $\bar{N}_{\rm H}^{\rm b}$, to each. This column density is computed on the filtered intensity map. The histogram of $\bar{N}_{\rm H}^{\rm b}$ for the filament sample is presented in Fig.~\ref{fig:3.3}. The 
number of filaments per $\bar{N}_{\rm H}^{\rm b}$ is represented by $N_{\rm F}$. 

\section{Interplay between the filament orientation and the magnetic field}\label{sec:hro}

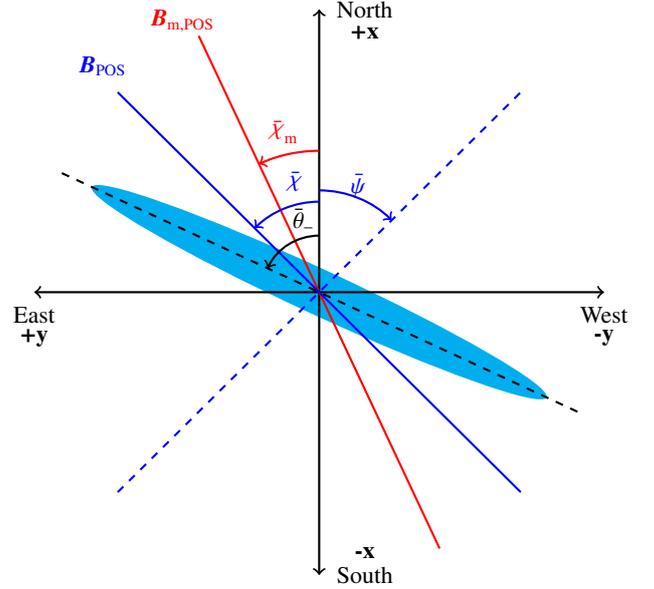
\begin{figure}
\centering
\begin{tikzpicture} [scale=1.5]
\fill[cyan] [rotate=-25] (0,0) ellipse (2.2 and 0.2);
\draw [<->][thick](-2.5,0) -- (2.5,0);
\draw [<->][thick](0,-2.5) --(0,2.5);
\node at (0.4,2.5) {North};
\node at (0.4,-2.5) {South};
\node at (2.5,-0.2) {West};
\node at (-2.5,-0.2) {East};
\node at (0.4,2.3) {\textbf{+x}};
\node at (0.4,-2.3) {\textbf{-x}};
\node at (2.5,-0.4) {\textbf{-y}};
\node at (-2.5,-0.4) {\textbf{+y}};
\draw[red] [thick][rotate=25] (0,-2.5) --(0,2.5);
\node[red] at (-1.2,2.4) {\bmean};
\path[red] (0,0)++(102:1.45cm)node{\anggmf};
\draw[red,->] [thick](0,1.25)arc(90:115:1.25cm);
\path[blue] (0,0)++(70:1.0cm)node{$\bar{\psi}$};
\draw[blue,->] [thick](0,0.9)arc(90:45:0.9cm);
\draw[blue] [thick][rotate=45] (0,-2.5) --(0,2.5);
\node[blue] at (-1.9,2.0) {\bpos};
\path[blue] (0,0)++(102:1.00cm)node{\angloc};
\draw[blue,->] [thick](0,0.8)arc(90:135:0.8cm);
\draw[blue,dashed] [thick][rotate=-45] (0,-2.5) --(0,2.5);
\draw[black,dashed] [thick][rotate=65] (0,-2.5) --(0,2.5);
\draw[black,->] [thick](0,0.5)arc(90:155:0.5cm);
\path (0,0)++(102:0.65cm)node{\hessangle};
\end{tikzpicture}
\caption{Sketch of the mean orientation angle of the filament (\hessangle), the magnetic field (\angloc), the polarization angle ($\bar{\psi}$), and the large-scale GMF (\anggmf) along the filament. All the angles are defined 
with respect to the GN and follow the IAU convention.}
\label{fig:4.1}
\end{figure}

In this section, we study the orientations of matter structures and \bpos\ in our filament sample (Sect.~\ref{sec:algorithm}). The orientation angle of \bpos\ is derived from 
the observed Stokes \Q\ and \U\ maps using Eqs.~\eqref{eq:2.2} and~\eqref{eq:2.3}. We also consider the orientation angle ($\rchi_{\rm m}$) of \bmean, as estimated from starlight polarization observations  
\citep{Heiles:1996} and pulsar rotation measures \citep{Rand:1994,Han:1999}. We compare these three orientations, as represented in Fig.~\ref{fig:4.1}. Our analysis follows \citet{planck2014-XXXII}, which 
used a set of pixels representing approximately 4\,\% of the sky at low and intermediate Galactic latitudes.  Only 25\,\% of the pixels in our current filament sample were considered in this earlier study.

\subsection{Relative orientation of the filaments and the magnetic field} \label{sec:4.1}

We study the angle difference between the orientations of the filaments and \bpos\ in our sample.  First we associate one POS orientation angle with each of the filaments with respect to the GN.  By 
construction, due to our selection criteria on the angles, the filaments are fairly straight and, hence, they may be described with a single orientation angle. Given one filament, we measure the mean orientation angle, 
\hessangle, over the $n$ pixels that belongs to it. We make use of the pseudo-vector field with unit length computed from the values of \mtheta\ for each pixel. This pseudo-vector has  components
$Q _{-} = \cos\ 2\mtheta $ and $U _{-} =  - \sin\ 2\mtheta $ (following the \healpix\ convention for the $Q _{-}$ and $U _{-}$ components). The mean POS orientation angle \hessangle\ of the filament is obtained by first 
averaging $Q _{-}$ and $U _{-}$ over all $n$ pixels and then calculating the position angle of this averaged pseudo-vector. It is given by

\begin{equation}
\hessangle= 0.5 \times \text{atan2} \left ( - \frac{1}{n} \sum_{i=1}^{n} U _{-} \ , \frac{1}{n}\sum_{i=1}^{n} Q _{-} \right)  \label{eq:4.1} \ .
\end{equation} 
If we rotate the Stokes \Q\ and \U\ maps by $\hessangle$, i.e., into the frame where the axis of the filament is in the North-South direction, the rotated \Q\ and \U\ can be written as
\begin{align}
Q '_{353}& = \Q \cos\ 2\hessangle\ - \U \sin\ 2\hessangle   \label{eq:4.2}\ , \\
U '_{353}& = \Q  \sin\ 2\hessangle\ + \U \cos\ 2\hessangle  \label{eq:4.3}\ .
\end{align}
Combining Eqs.~\eqref{eq:4.2} and~\eqref{eq:4.3} with Eqs.~\eqref{eq:2.1} and~\eqref{eq:2.2}, we get
\begin{align}
Q'_{353} & =  \phantom{-} \  P_{353} \cos\ 2(\psi - \hessangle) =  - \ P_{353} \cos\ 2(\rchi - \hessangle) \ , \\
U'_{353} & =  - \ P_{353} \sin\ 2(\psi - \hessangle) =   \phantom{-} \ P_{353} \sin\ 2(\rchi - \hessangle) \ ,
\end{align} 
where the orientation angle $\rchi$ is defined in Eq.~\eqref{eq:2.3}. Similar to the computation of \hessangle, we average $Q'_{353}$ and $U'_{353}$ over  all $n$ pixels and then calculate the position angle of this averaged pseudo-vector
\begin{align}
\bar{Q}'_{353} &=   \frac{1}{n}\sum_{i=1}^{n} Q'_{353} \equiv - \   \bar{P}_{353} \cos\ 2 \Delta_{\angloc - \hessangle} \label{eq:4.4}\ , \\
\bar{U}'_{353} &=   \frac{1}{n}\sum_{i=1}^{n} U'_{353} \equiv \phantom{-} \  \bar{P}_{353} \sin\ 2 \Delta_{\angloc - \hessangle}  \label{eq:4.5}\ ,
\end{align}
where
\begin{align}
\Delta_{\angloc - \hessangle}  & = 0.5 \times \text{atan2}  ( \bar{U}'_{353}, - \bar{Q}'_{353} ) \ ,  \label{eq:4.6}\\
\bar{P}_{353} & = \sqrt{{\bar{Q}_{353}}^{'2} +  {\bar{U}_{353}}^{'2}} \ .
\end{align}
The angle difference $\Delta_{\angloc - \hessangle}$ measures the weighted mean of the angle difference per pixel between the orientations of the given filament and \bpos. The index \angloc\ refers to the mean orientation angle of \bpos\  along the filament. Note that we directly measure the angle difference between the filament  and \bpos, without computing \angloc\ for each filament.

\begin{figure}[h!]
\begin{tabular}{c}
\includegraphics[width=8.8cm]{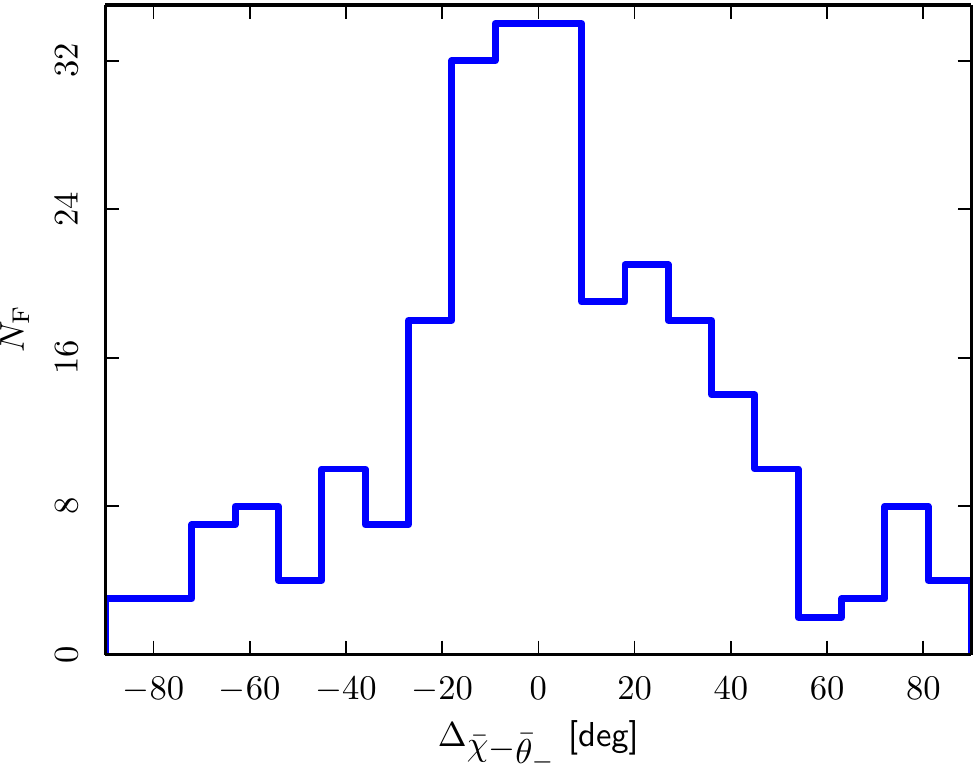} \\
\includegraphics[width=8.8cm]{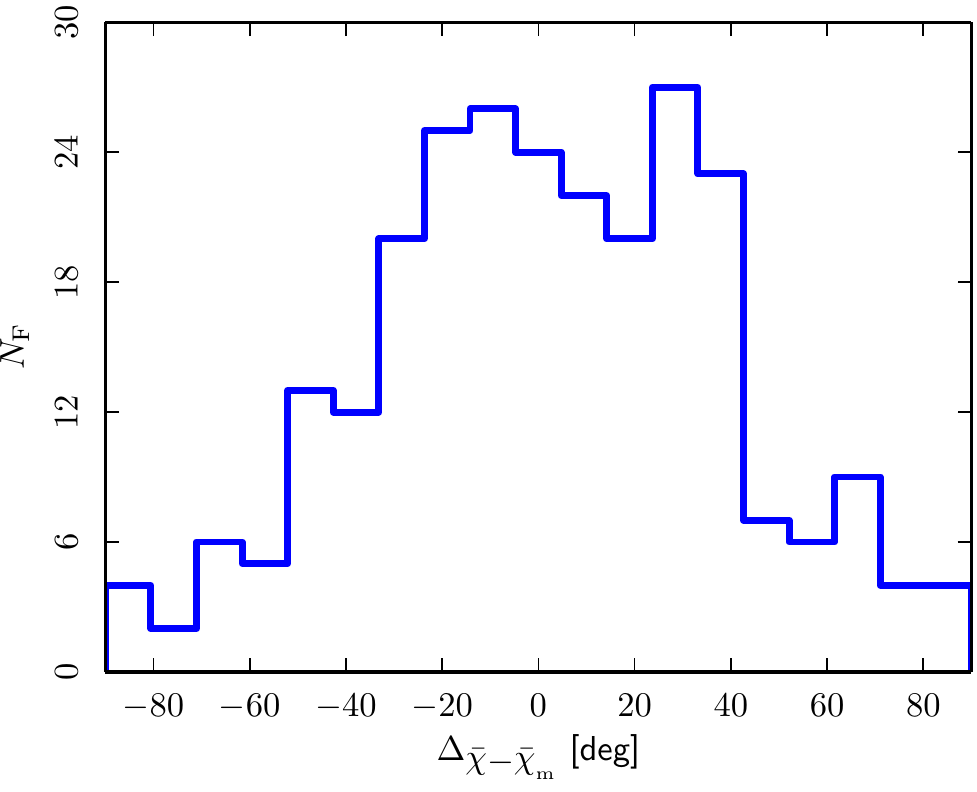} \\
\includegraphics[width=8.8cm]{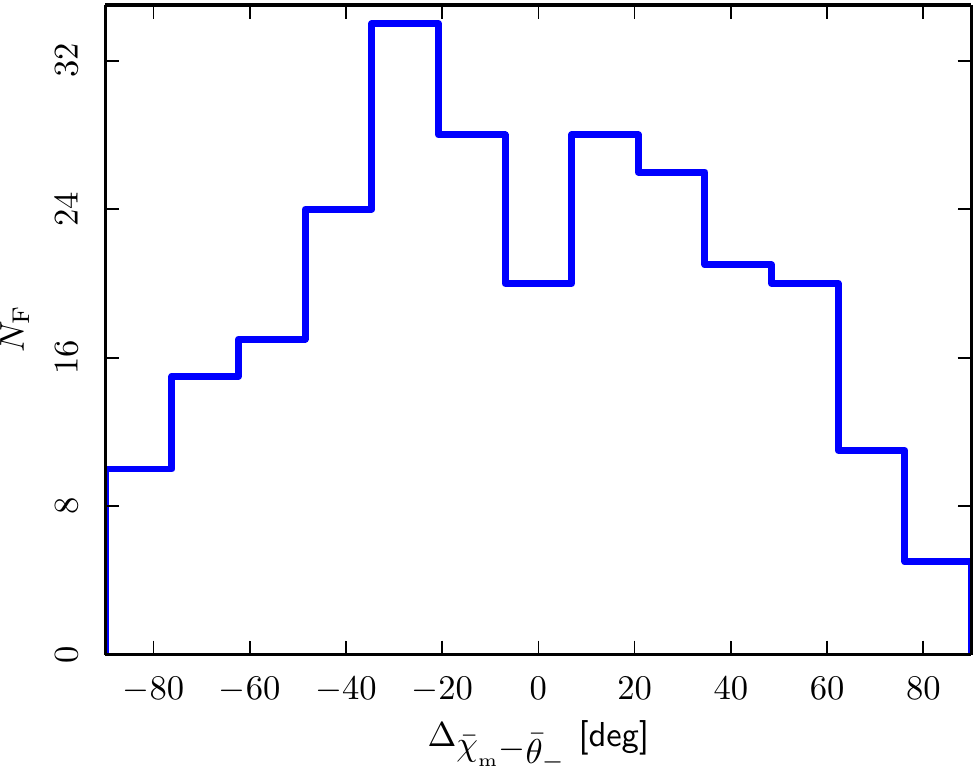} 
\end{tabular}
\caption{\textit{Upper panel}: HRO between the filaments and \bpos. \textit{Middle panel}: HRO between \bpos\ and \bmean.   \textit{Lower panel}:  HRO between the  filaments and \bmean. }
\label{fig:4.2}
\end{figure}

The histogram of relative orientation (HRO) between the filament and \bpos\ for our filament sample is presented in the upper panel of Fig.~\ref{fig:4.2}.  The mean value of the histogram is $2\pdeg3$ computed using the 
equivalent of Eq.~\eqref{eq:4.1}. Our histogram agrees with the pixel-by-pixel analysis at intermediate and low Galactic latitudes presented in \citet{planck2014-XXXII}. Like in this earlier study, we find that the filaments 
are statistically aligned with \bpos. A similar alignment between the filaments in the intensity map and \bpos\ has been reported for synchrotron emission observed by \wmap\  at 23 GHz \citep{Vidal:2014}.

To quantify the shape of the histogram of $\Delta_{\angloc - \hessangle}$, we fit it with a Gaussian plus a constant. The Gaussian has a $1\,\sigma$ dispersion of 19\deg. The constant may be accounted for by the projection of the magnetic field and filament orientations on the POS as discussed  in \citet{planck2014-XXXII}.

\begin{figure*}
\begin{tabular}{c}
\includegraphics[width=18cm]{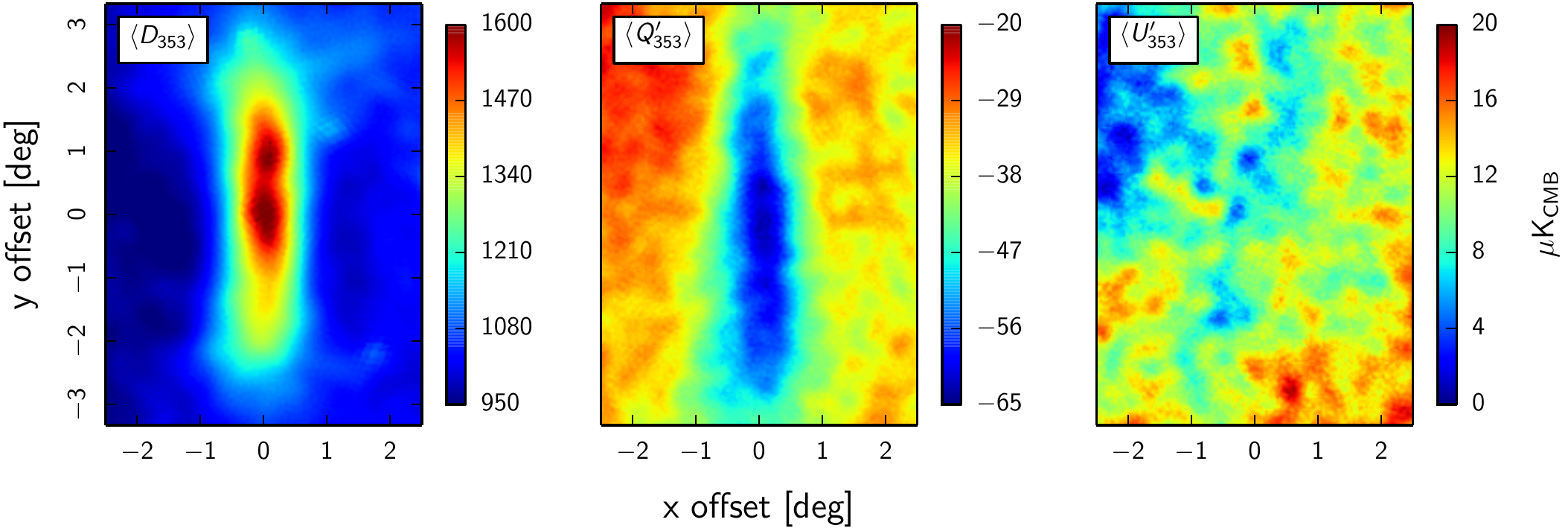} 
\end{tabular}
\caption{Mean images of  the \planck\   $\langle \Dmodel  \rangle$ ,  $\langle \bQ  \rangle$,  and  $\langle \bU \rangle$   maps over the \mlambda\ filaments  at 15\arcm\ resolution. }
\label{fig:5.1}
\end{figure*}

\subsection{Relative orientation of the magnetic field and the large-scale Galactic magnetic field} \label{sec:4.2}

Here, we compare the orientation of \bpos\ on the filaments with that of \bmean. \citet{Heiles:1996} derived the orientation of \bmean\ pointing towards $l_0=82\pdeg8 \pm 4\pdeg1$ and $b_0=0\pdeg4 \pm 0\pdeg5$ 
from the polarization pseudo-vectors of stars more distant than 500 parsecs. Slightly different $l_0$ values have been reported in other studies. From the rotation measures of nearby pulsars within a few hundred parsecs of the Sun, 
\citet{Rand:1994} found the direction of \bmean\  pointing towards $l_0=88\deg \pm 5 \deg$. In another study of pulsar rotation measures, \citet{Han:1999} derived the direction of \bmean\  as $l_0 \simeq 82$\deg. 
These two studies do not report values for $b_0$, which is assumed to be zero. Based on these observations, we assume that the mean orientation of \bmean\  in the solar neighbourhood is $l_0=84\deg \pm10$\deg\ and 
$b_0=0\deg \pm 10\deg$, with the same uncertainty on $l_0$ and $b_0$. 

We construct a pseudo-vector field with unit length based on the uniform orientation of \bmean. This pseudo-vector has components: $Q _{\rm m} = \cos\ 2\psi_{\rm m} = \cos\ 2(\rchi_{\rm m}- \pi/2)$ and 
$U _{\rm m} = - \sin\ 2\psi_{\rm m} = - \sin\ 2(\rchi_{\rm m}- \pi/2)$ (following the \healpix\ convention for the $Q _{\rm m}$ and $U _{\rm m}$ maps), where $\psi_{\rm m}$ is the polarization angle of \bmean. The procedure 
to go from the uniform \bmean\ pointing towards ($l_0, b_0$) to $\psi_{\rm m}$ is detailed by \citet{Heiles:1996}. The mean orientation angle (\anggmf) of \bmean\ for each filament  is obtained by first averaging 
$Q _{\rm m}$ and $U _{\rm m}$ over all $n$ pixels within a filament and then calculating the position angle of this averaged pseudo-vector. We compute the angle difference, $\Delta_{\angloc-\anggmf}$, between the 
orientations of \bpos\ and \bmean\ on the filament in a similar manner to the method described in Sect.~\ref{sec:4.1}. 

The HRO between \bpos\ and \bmean\  for our filament sample is presented in the  middle panel of Fig.~\ref{fig:4.2}. The mean value of the histogram is $1\pdeg3 \pm 3\pdeg7$ where the uncertainty  is computed by 
changing the mean orientation of \bmean\ within its quoted uncertainties. This HRO has a larger dispersion than that between  the filaments and \bpos\  shown in the upper panel of Fig.~\ref{fig:4.2}.  To quantify the 
shape of the histogram of $\Delta_{\angloc - \anggmf}$, we fit it with a Gaussian plus a constant. The Gaussian has a $1\,\sigma$ dispersion of 36\deg. 

We conclude that \bpos\ of our filament sample is statistically aligned with \bmean. \citet{planck2014-XXXII} reports a similar correlation, for the low and intermediate Galactic latitudes, when comparing the polarization 
measured on the filaments with their background polarization maps. The scatter measured by the HRO may be interpreted considering both the turbulent 
component of the magnetic field and projection effects. 


\subsection{Relative orientation of the large-scale Galactic magnetic field and the filaments} \label{sec:4.3}

We combine the results obtained in Sects.~\ref{sec:4.1} and~\ref{sec:4.2} to assess statistically the orientation of \bmean\ in the solar neighbourhood with respect to the filaments. \bpos\ is statistically 
aligned with the filaments in our sample and with \bmean. From both results, one would intuitively expect \bmean\ to be statistically aligned  with the filaments. To test this expectation, for each filament, 
we compute the angle difference, $\Delta_{\anggmf-\hessangle}$, between the orientations of \bmean\ and the filament. The angle difference $\Delta_{\anggmf-\hessangle}$ is computed in a similar manner to the method described 
in Sect.~\ref{sec:4.1}.

The HRO between the filament and \bmean\   for our filament sample is presented in the  lower panel of Fig.~\ref{fig:4.2}. A correlation between the orientation angles of the filament and \bmean\ is present, but the HRO shows more scatter than the HRO between the filaments and \bpos\ and that between \bpos\ and \bmean. The histogram has a mean value of $-3\pdeg1 \pm 2\pdeg6$. To quantify the shape of the histogram of 
$\Delta_{\angloc - \anggmf}$, we fit with a Gaussian plus a constant. The Gaussian has a $1\,\sigma$ dispersion of 54\deg. \cite{planck2014-XXXII} reported a similar loss of correlation when comparing the orientations of 
the filaments with that of \bpos\ derived from their local background polarization maps.

\section{Mean polarization properties of filaments}\label{sec:pol_frac}

In this section, we present stacked images of the filaments in Stokes $I$, $Q$, and $U$, after rotation to align the filaments and to compute 
$Q$ and $U$ with respect to their orientation. The images are used to compute the average polarization fraction of our sample of filaments. 

\subsection{Stacking filaments}\label{sec:5.1}

Over the high-latitude sky, the signal-to-noise ratio of the 353\,GHz \planck\ polarization maps  is low and it is not possible to measure the polarization fraction of individual dust intensity filaments in our sample. In order to increase the signal-to-noise ratio, we therefore stack images of the 259 filaments and their surroundings.

For each filament in the sample, using the ``\gnomview" routine of \healpix, we extract from the \Planck\ maps  a  local, flat-sky, image ($7\deg \times 5\deg$ patch) centred on the filament centre and rotated by \hessangle\ 
in the clockwise direction to align the filament in the North-South direction. We stack the images of the filaments in \Dmodel,  \bQ, and \bU\ (as defined in Eqs.~\ref{eq:4.2} and~\ref{eq:4.3}) after aligning all the maps in the 
North-South direction. We produce mean stacked images, denoted with angle brackets $\langle .. \rangle$, by dividing the  sum of the individual images by the total number of filaments in our sample; they are presented in Fig.~\ref{fig:5.1}. 
The $1\,\sigma$ errorbar both on the $\langle \bQ \rangle$ and $\langle \bU \rangle$ images is $1.3$\, \mukcmb, as computed from the difference of two polarization HM maps. All the features presented in Fig.~\ref{fig:5.1} are significant compared to the data systematics and statistical noise.
The average filament appears as a negative feature with respect to the background in the $\langle \bQ \rangle$ image and is not seen in the $\langle \bU \rangle$ image. This result is a 
direct consequence of the alignment between the filaments and \bpos\ \citep[Sect.~\ref{sec:4.1} and ][]{Zaldarriaga:2001}. The background in both  the $\langle \bQ \rangle$ and $\langle \bU \rangle$ images is rather homogeneous. This reflects the smoothness of \bpos\ within the $7\deg \times 5\deg$ patches. 

We perform a null test to assess the significance of the stacking of filaments. This test is made by stacking 259 randomly chosen $7\deg \times 5\deg$ patches in the high-latitude sky. Each patch is rotated in the clockwise direction, with the orientation angle \mtheta\ of the central pixel.  The images of $\langle \Dmodel  \rangle$, $\langle \bQ  \rangle$, and $\langle \bU \rangle$ for random patches are consistent with noise.
The amplitude of $\langle \bQ  \rangle$ and $\langle \bU \rangle$ images is comparable to that of the difference between stacked images obtained when applying the same analysis to each of the two polarization HM maps.
This confirms the hypothesis that the filaments detected in Fig.~\ref{fig:5.1} are indeed real and are rotated with a well-determined angle \hessangle.

\subsection{Polarization fraction}\label{sec:5.2}

\begin{figure}
\begin{tabular}{c}
\includegraphics[width=8.8cm]{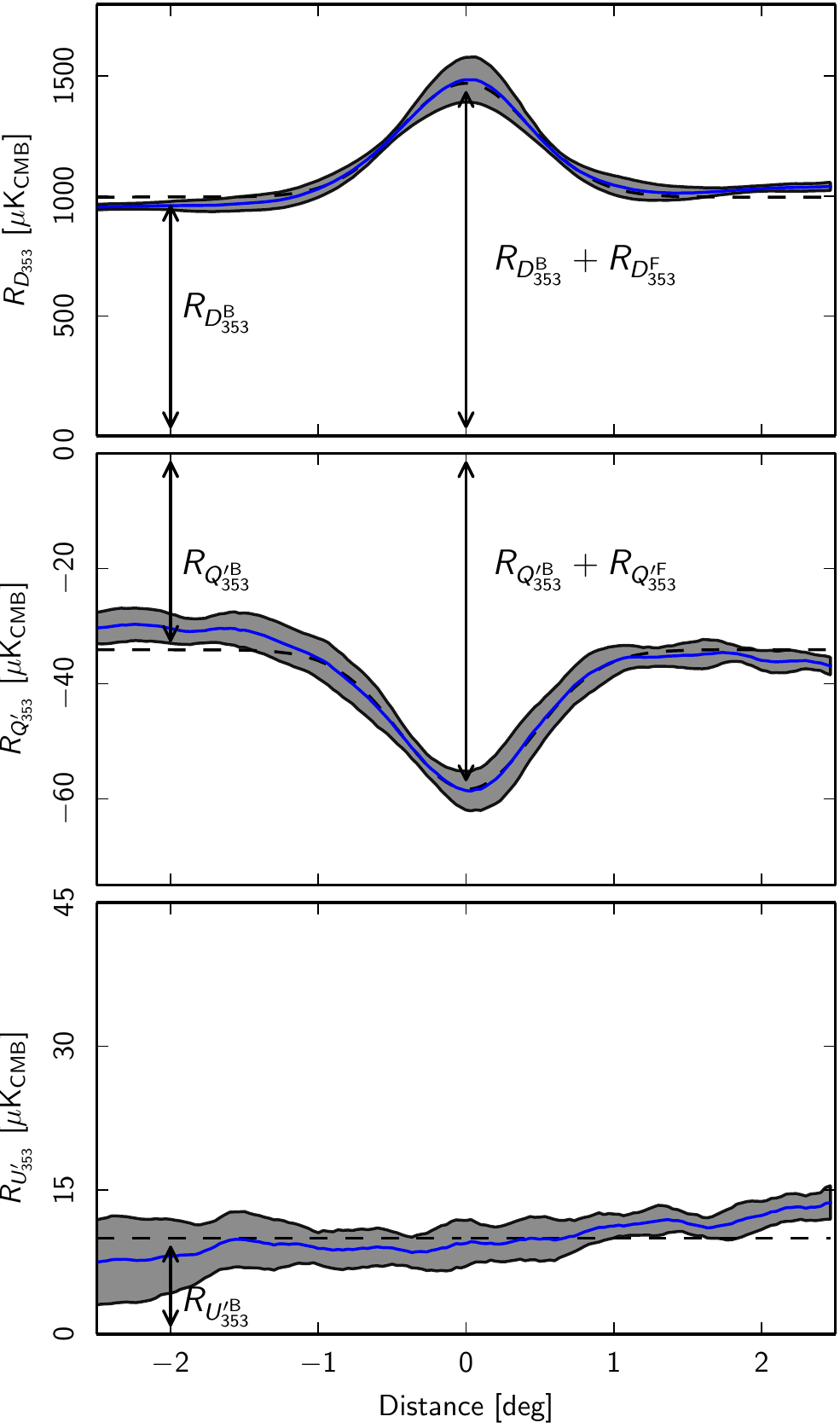}
\end{tabular}
\caption{Radial profiles of the mean stacked \planck\ \Dmodel, \bQ, and \bU\ images as functions of distance from the centre of the filament (blue line). The grey shaded region shows the $1\,\sigma$ dispersion from the data values at a given radial distance from the filament axis that we average. The dashed line is the Gaussian fit to the filament profile plus a constant background emission.}
\label{fig:5.2}
\end{figure}

Instead of using individual pixels, we collapse the mean stacked images in the filament direction to draw the radial profiles ($R$) of the \Dmodel, \bQ, and \bU\ images,  which are presented in Fig.~\ref{fig:5.2}. The 
shaded area in Fig.~\ref{fig:5.2} represents the $1\,\sigma$ dispersion from the data values at a given radial distance from the filament axis that we average.  We clearly identify the profile of the filament on top of the constant background emission in the \Dmodel\ and \bQ\ radial 
profiles, while the radial profile of \bU\ is consistent with a constant background emission. The radial profiles of \Dmodel, \bQ, and \bU\ can be decomposed into the filament ($\rm F$) and the background  ($\rm B$) contributions as
\begin{align}
R_{\Dmodel}  & =  R_{\bFD} + R_{\bBD} \ , \\ 
R_{\bQ}  & =  R_{\bFQ} + R_{\bBQ} \ , \\ 
R_{\bU}  & = R_{\bFU} +  R_{\bBU}  \ .   
\end{align}
We fit the radial profiles of \Dmodel\ and \bQ\ in Fig.~\ref{fig:5.2} with a Gaussian profile for the filament emission plus a constant for the background emission. We find that the centre of the Gaussian profile is zero and that their $1\,\sigma$ dispersion is $27\arcm\pm1\arcm$ for both the \Dmodel\ and \bQ\ radial profiles.

Following Eqs.~\eqref{eq:4.4} and~\eqref{eq:4.5}, we can express the average Stokes \bQ\ and \bU\ for one given filament as
\begin{align}
\abFQ &= -  \ \bar{P}_{353}^{\rm F} \cos\ 2 \Delta_{\angloc - \hessangle}^{\rm F} = - \ \bar{p}^{\rm F} \ \bar{D}_{353}^{\rm F} \cos\ 2 \Delta_{\angloc - \hessangle}^{\rm F} \ , \label{eq:6.m1}\\
\abFU&=   \phantom{-}  \ \bar{P}_{353}^{\rm F} \sin\ 2 \Delta_{\angloc - \hessangle}^{\rm F} =  \phantom{-} \  \bar{p}^{\rm F} \  \bar{D}_{353}^{\rm F} \sin\ 2 \Delta_{\angloc - \hessangle}^{\rm F}\ , \label{eq:6.0}
\end{align}
where $\bar{D}_{353}^{\rm F}$ and $\bar{P}_{353}^{\rm F}$ are the average specific intensity and polarization intensity of the filament. The superscript $\rm F$ represents the contribution from the filament only. The polarization fraction ($\bar{p}^{\rm F}$) of a filament is defined by 
\begin{equation}
\bar{p}^{\rm F}= \frac{\bar{P}_{353}^{\rm F}} {\bar{D}_{353}^{\rm F}}. 
\end{equation}
From Fig.~\ref{fig:3.3}, we know that most filaments in our sample have comparable column densities and hence  roughly the same $\bar{D}_{353}^{\rm F}$. The mean stacked Stokes \abFQ\ and \abFU\ values for all the filaments can be approximated as
\begin{align}
\langle \abFQ \rangle& \simeq -  \langle \bar{p}^{\rm F}  \cos\ 2 \Delta_{\angloc - \hessangle}^{\rm F}  \rangle \langle \bar{D}_{353}^{\rm F} \rangle \simeq  - \langle  \bar{p}^{\rm F} \rangle \langle \cos\ 2 \Delta_{\angloc - \hessangle}^{\rm F}  \rangle \langle \bar{D}_{353}^{\rm F} \rangle, \label{eq:6.1}\\
\langle \abFU \rangle&\simeq  \phantom{-}    \langle \bar{p}^{\rm F}  \sin\ 2 \Delta_{\angloc - \hessangle}^{\rm F}  \rangle \langle \bar{D}_{353}^{\rm F} \rangle \simeq  \phantom{-}  \langle  \bar{p}^{\rm F} \rangle \langle \sin\ 2 \Delta_{\angloc - \hessangle}^{\rm F}  \rangle \langle \bar{D}_{353}^{\rm F} \rangle , \label{eq:6.2}
\end{align}
where $\langle  \bar{p}^{\rm F} \rangle$ is the mean polarization fraction of our filament sample.  For the radial profiles of the filament emission, we have 
\begin{align}
R_{\bFQ} & \simeq  - \langle  \bar{p}^{\rm F} \rangle \langle \cos\ 2 \Delta_{\angloc - \hessangle}^{\rm F}  \rangle R_{\Dmodel^{\rm F}} \ , \\
R_{\bFU} & \simeq  \phantom{-}  \langle  \bar{p}^{\rm F} \rangle \langle \sin\ 2 \Delta_{\angloc - \hessangle}^{\rm F}  \rangle R_{\Dmodel^{\rm F}} \ .
\end{align}  
The angle difference between the filaments and \bpos\ (Eq.~\ref{eq:4.6}) is used for $\Delta_{\angloc - \hessangle}^{\rm F}$. The histogram of $\Delta_{\angloc - \hessangle}^{\rm F}$ is roughly symmetric around 0\deg, implying  $R_{\bFU} \ll R_{\bFQ}$. 

Similarly, the radial profiles from the background emission can be written as
\begin{align}
R_{Q^{\rm 'B}_{\rm 353}} & \simeq  - \langle  \bar{p}^{\rm B} \rangle \langle \cos\ 2 \Delta_{\angloc - \hessangle}^{\rm B}  \rangle R_{\Dmodel^{\rm B}} \ , \label{eq:6.3} \\
R_{U^{\rm 'B}_{\rm 353}} &\simeq  \phantom{-}  \langle  \bar{p}^{\rm B} \rangle \langle \sin\ 2 \Delta_{\angloc - \hessangle}^{\rm B}  \rangle R_{\Dmodel^{\rm B}}  \label{eq:6.4}\ ,
\end{align}  
where $\langle  \bar{p}^{\rm B} \rangle$ is the average polarization fraction of the background emission and $\Delta_{\angloc - \hessangle}^{\rm B}$ is the angle difference between a given filament and \bpos\ from its local background polarization.

\begin{figure}
\begin{tabular}{c}
\includegraphics[width=8.8cm]{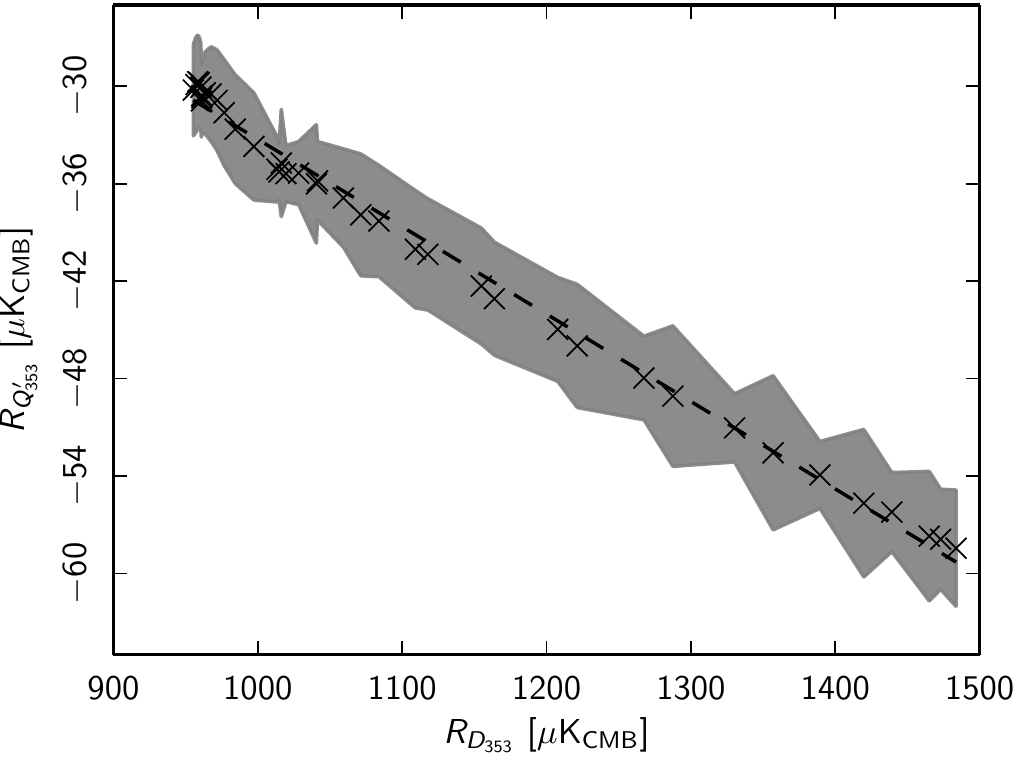}
\end{tabular}
\caption{Linear fit (dashed line) to the correlation between the mean radial profiles of the stacked \bQ\ and \Dmodel\ images. The grey shaded region shows the $1\,\sigma$ dispersion from the data values at a given radial distance from the filament axis that we average. }
\label{fig:5.3}
\end{figure}

The observed radial profile of the stacked \bQ\ image in Fig.~\ref{fig:5.2} can be written as
\begin{align}
&R_{\bQ}   =  R_{\bFQ}  +R_{Q^{\rm 'B}_{\rm 353}}      \nonumber \\
& \simeq    -  \langle \bar{p}^{\rm F} \rangle \langle \cos\ 2 \Delta_{\angloc - \hessangle}^{\rm F}\rangle  R_{\Dmodel^{\rm F}}   -   \langle \bar{p}^{\rm B}  \rangle \langle \cos\ 2 \Delta_{\angloc - \hessangle}^{\rm B}   \rangle  R_{\Dmodel^{\rm B}}   \nonumber\\
&=  -  \langle \bar{p}^{\rm F}  \rangle \langle \cos\ 2 \Delta_{\angloc - \hessangle}^{\rm F}  \rangle  R_{\Dmodel}  \nonumber  \\
& \hspace{2cm} + \left [ \langle \bar{p}^{\rm F}  \rangle \langle \cos\ 2 \Delta_{\angloc - \hessangle}^{\rm F}  \rangle - \langle \bar{p}^{\rm B} \rangle  \langle \cos\ 2 \Delta_{\angloc - \hessangle}^{\rm B}  \rangle \right ]  R_{\Dmodel^{\rm B}}  \nonumber \\
&=  a \  R_{\Dmodel} + b  \ ,
\end{align}
where $a=-\langle \bar{p}^{\rm F}  \rangle \langle \cos\ 2 \Delta_{\angloc - \hessangle}^{\rm F}  \rangle$ is the scaling parameter and $b=\left [  \langle \bar{p}^{\rm F}  \rangle \langle \cos\ 2 \Delta_{\angloc - \hessangle}^{\rm F}  \rangle -  \langle \bar{p}^{\rm B} \rangle  \langle \cos\ 2 \Delta_{\angloc - \hessangle}^{\rm B}  \rangle \right]  R_{\Dmodel^{\rm B} }$ is the offset of the linear fit between the $R_{\bQ}$ and $R_{\Dmodel}$ profiles.
The linear fit between $R_{\bQ}$ and $R_{\Dmodel}$  is shown in Fig.~\ref{fig:5.3}. The best-fit  parameter values from the linear fit are
\begin{align}
\langle \bar{p}^{\rm F}  \rangle \langle \cos\ 2 \Delta_{\angloc - \hessangle}^{\rm F}  \rangle &= 5.3\,\% \ ,  \label{eq:6.5}\\
\left [\langle \bar{p}^{\rm F}  \rangle \langle \cos\ 2 \Delta_{\angloc - \hessangle}^{\rm F}  \rangle\ - \langle \bar{p}^{\rm B}  \rangle \langle \cos\ 2 \Delta_{\angloc - \hessangle}^{\rm B}  \rangle \right ]  R _{\Dmodel^{\rm B}} &= 19.8\, \mukcmb \ . \label{eq:6.6}
\end{align}

In our estimate of $\langle \bar{p}^{\rm F} \rangle \langle \cos\ 2 \Delta_{\angloc - \hessangle}^{\rm F}  \rangle$, the noise bias is negligible, because the noise averages out in the stacking of the \bQ\ and \bU\ maps. The HRO between the filament and \bpos\  (upper panel in Fig.~\ref{fig:4.2}) is used to 
compute $\langle \cos\ 2 \Delta_{\angloc - \hessangle}^{\rm F}\rangle$, which is 0.48.  Taking this factor into account, the mean polarization fraction of our filament sample is 
\begin{equation}
\langle \bar{p}^{\rm F} \rangle =11\,\%. 
\end{equation}
We have computed $\langle  \bar{p}^{\rm F} \rangle$ using the two independent subsets of the \planck\ data (HM maps). For both data sets, we find the same mean value of 11\,\%.  This is expected, because the 
HRO between the filaments and \bpos\ is determined with high accuracy (Appendix~\ref{sec:angle_errors}). It shows that the measurement error on $\langle \bar{p}^{\rm F}\rangle $ is small. 
Obviously $\langle  \bar{p}^{\rm F}\rangle $ may be different for another set of filaments, because it depends on the angles of the filaments with respect to the plane of the sky. 

The mean polarization fraction of the filaments determined from the filament sample is smaller than the maximum degree of polarization reported in \citet{planck2014-XIX}, which is $p_{\rm max}=19.6$\,\%. This result 
suggests that there is some depolarization due to changes in the magnetic field orientation within the filaments \citep{planck2014-XX,planck2014-XXXIII}.

The offset $b$ from the linear fit to the radial profiles of the \bQ\ and \Dmodel\ images is positive. This means that
\begin{align}
\langle \bar{p}^{\rm B} \rangle  \langle \cos\ 2 \Delta_{\angloc - \hessangle}^{\rm B}  \rangle &<  \langle\bar{p}^{\rm F}  \rangle \langle \cos\ 2 \Delta_{\angloc - \hessangle}^{\rm F}  \rangle\  \label{eq:6.7}
\end{align} 
The distribution of $\Delta_{\angloc - \hessangle}^{\rm B}$ for our filament sample is needed to compute $\langle  \bar{p}^{\rm B} \rangle$. We make a simple approximation using the HRO between the filament and 
\bmean\ (lower panel in Fig.~\ref{fig:4.2}) as a proxy for  $\Delta_{\angloc - \hessangle}^{\rm B}$. We then find $\langle \cos 2\Delta_{\angloc - \hessangle}^{\rm B} \rangle =0.25\pm0.14$ for our filament sample. 
Combining Eqs.~\eqref{eq:6.5} and~\eqref{eq:6.7}, we put a upper limit on the mean polarization fraction of the background emission as
\begin{equation}
\langle  \bar{p}^{\rm B} \rangle < 21\,\% _{-7\,\%} ^{+27\,\%} \ . 
 \end{equation} 
This upper limit on $\langle  \bar{p}^{\rm B} \rangle$ depends on the mean orientation of \bgmf\ and it ranges from 48\,\% to 14\,\% for the 10\deg\ uncertainty on $l_0$ and $b_0$ (Sect.~\ref{sec:4.2}). The profile of $R_{U^{\rm 'B}_{\rm 353}}$ in the lower 
panel of Fig.~\ref{fig:5.2} is roughly constant and positive. 
This result indicates that \bpos\ is smooth within the $7\deg \times 5\deg$ patches and $\langle \sin 2\Delta_{\angloc- \hessangle}^{\rm B}\rangle$ is positive (Eq.~\ref{eq:6.4}), which follows if the distribution of $\Delta_{\angloc- \hessangle}^{\rm B}$ is not symmetric 
with respect to 0\deg. We point out that the histogram in the lower panel in Fig.~\ref{fig:4.2} may not be used to estimate the sign of $\langle \sin 2\Delta_{\angloc- \hessangle}^{\rm B} \rangle$ because the uncertainty on $l_0$ and $b_0$ is too large.

\section{\EB\ asymmetry}\label{sec:eb_asymmetry}

In this section, we quantify the \EB\ asymmetry of dust polarization, over the $\ell$ range 30 to 300, for our filament sample, and relate it 
to the relative orientation between filaments and \bpos. 
First, we present the stacked images of filtered maps that we use to quantify the \EB\ asymmetry. Second, we compute the contribution of the pixels used in the stacking to the variance in the high-latitude sky. Last, we present an analytical approximation that relates the HRO between the filaments and \bpos\ to the \EB\ asymmetry. 

\begin{figure*}
\includegraphics[width=17.5cm]{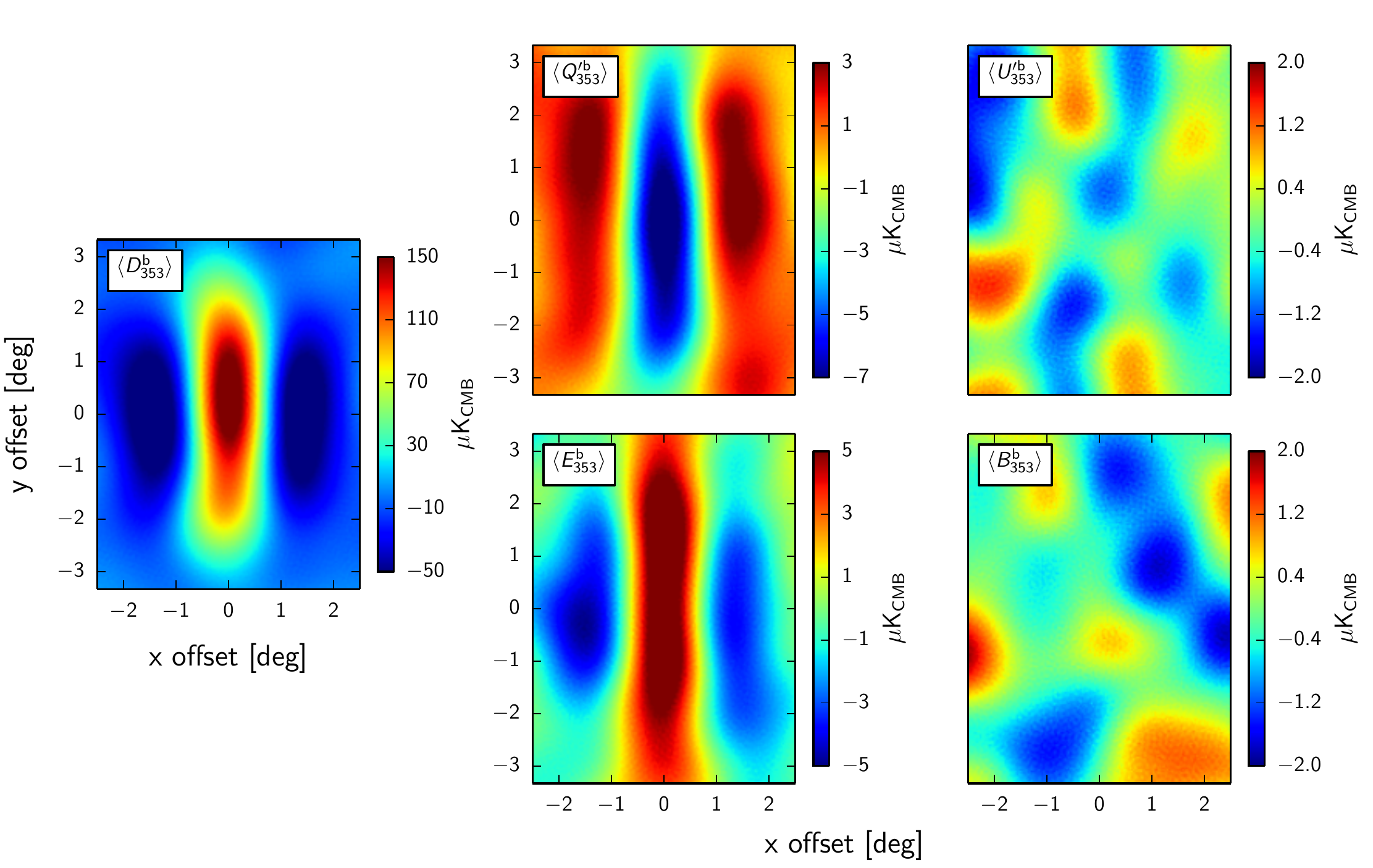}  
\caption{Mean images of  the \planck\ $\langle \fDmodel \rangle$, $\langle \fQ \rangle$,  $\langle \fU \rangle$, $\langle  \fE \rangle$, and $\langle \fB \rangle$  maps over the \mlambda\ filaments. }
\label{fig:6.1}
\end{figure*}

\begin{figure*}
\includegraphics[width=17.5cm]{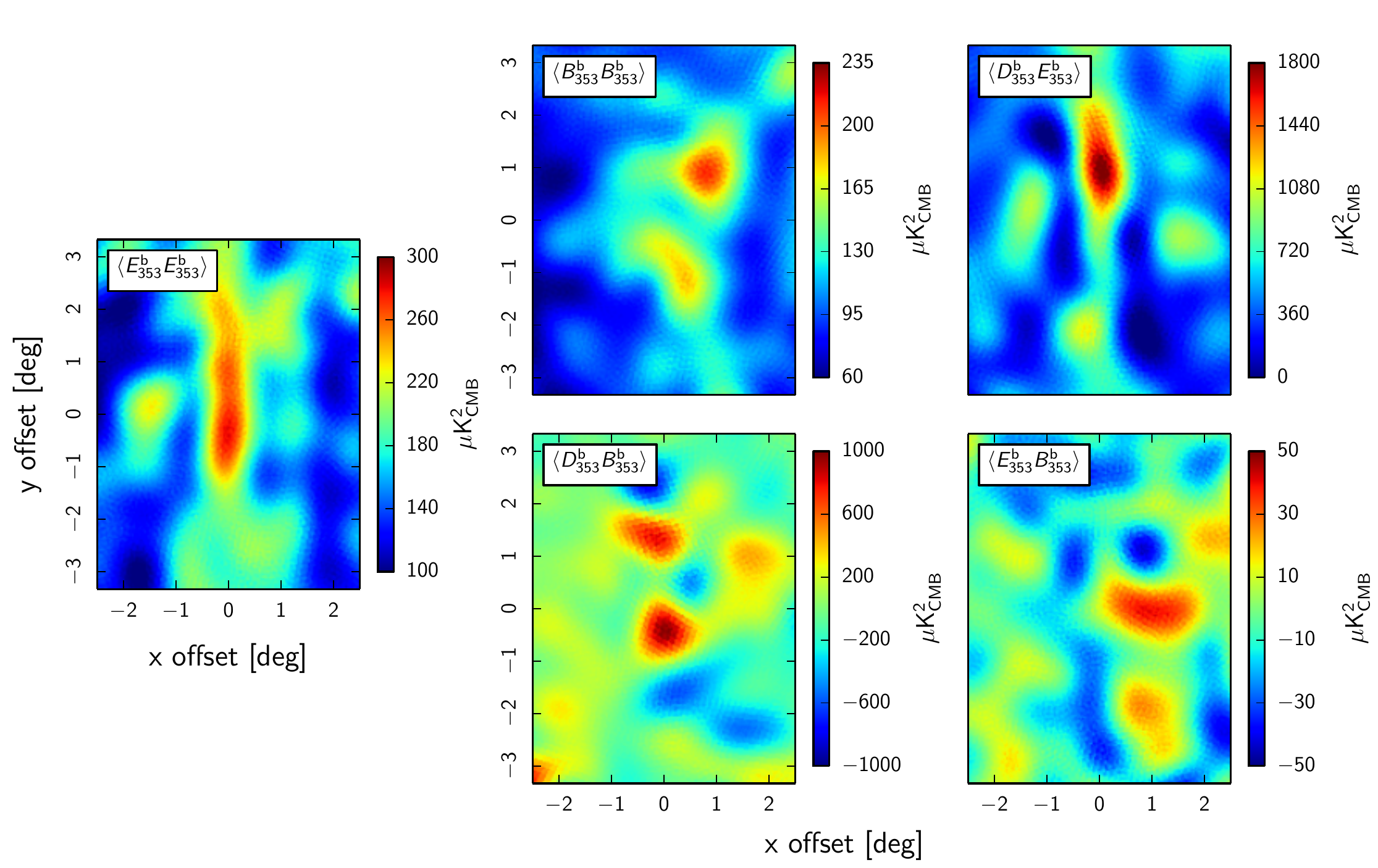}  
\caption{Mean images of  the  \planck\ $\langle \fE \fE \rangle$, $\langle \fB \fB \rangle$,  $\langle \fDmodel  \fE \rangle$,  $\langle \fDmodel  \fB \rangle$, and  $\langle \fE  \fB \rangle$  maps over the \mlambda\ filaments. }
\label{fig:6.2}
\end{figure*}

\subsection{Stacked images of filtered maps}\label{sec:6.1}

Here we use the \planck\ \fDmodel, $Q_{\rm 353}^{\rm b}$, $U_{\rm 353}^{\rm b}$, \fE, and \fB\ maps to analyse the $\ell$ range, between 30 and 300, over which we identify the filaments in the dust intensity map. Our choice of $\ell$ range has a large overlap with the angular scales, $40 < \ell < 600$, where the \EB\ asymmetry has been measured with the power spectra of dust polarisation. We stack all the bandpass-filtered maps, rotated by the mean orientation angle \hessangle\ of the filament, as described in Sect.~\ref{sec:5.1}. The \fQ\ and \fU\ maps are the filtered Stokes $Q_{\rm 353}$ and $U_{\rm 353}$ maps computed with respect to the axis of the filament. The mean stacked images of the bandpass-filtered maps are presented in  Fig.~\ref{fig:6.1}. The sidelobes that appear in the images of $\langle \fDmodel \rangle$, $\langle \fQ \rangle$, and $\langle \fE \rangle$ of Fig.~\ref{fig:6.1} on both sides of the filament centre are coming from the filtering. The $1\,\sigma$ errorbar on the $\langle \fQ \rangle$, $\langle \fU \rangle$, $\langle \fE \rangle$, and $\langle \fB \rangle$ images is $0.2$\, \mukcmb, as computed from the difference of two polarization HM maps. The filaments appear as a negative feature in the $\langle \fQ \rangle$ image and a positive feature in that of $\langle \fE \rangle$. The $\langle \fU \rangle$ and $\langle \fB \rangle$ images are consistent with a mean value zero. 

Next, we stack the products of two quantities, i.e.,  $\langle \fDmodel \fE \rangle$, $\langle \fDmodel \fB \rangle$, $\langle \fE \fE \rangle$, $\langle \fE \fB \rangle$, and $\langle \fB \fB \rangle$. We follow the same methodology as discussed in Sect.~\ref{sec:5.1} with oriented stacking at the centre of the filament. To avoid a noise bias in the square quantities, $\langle \fE \fE \rangle$ and $\langle \fB \fB \rangle$, we compute the cross-product of the two HalfMission maps (HM1 and HM2). For other quantities, $\langle \fDmodel \fE \rangle$, $\langle\fDmodel \fB  \rangle$, and  $\langle\fE \fB  \rangle$, we use the full-mission maps. The images produced by stacking the products are presented in Fig.~\ref{fig:6.2}.  Note that  the maps shown in Figs.~\ref{fig:6.1} and~\ref{fig:6.2} are not used for data analysis, but still we comment on these images. We computed the mean stacked images from each of the two polarization DS or HR maps and all the features presented in Fig.~\ref{fig:6.2} appear for both independent subset of \planck\ data. In the idealized description of the filaments in \citet{Zaldarriaga:2001}, we will expect the average filament to appear similarly in the $EE$, $BB$, and $TE$ maps. The differences between these three images shows that the reality of the dust sky is more complex than the idealized model.

\subsection{Measured \EB\ asymmetry}\label{sec:6.2}

We measure the \EB\ asymmetry at the filtering scale over the high-latitude ({\rm HL}) region. We compute the variances ($V$) of the \fE\ and \fB\ maps using the relations
\begin{align}
V^{EE} \ ({\rm HL}) &=  \frac{1}{N_{\rm HL}} \sum_{i=1}^{N_{\rm HL}} E^{\rm b}_{353,\rm HM1} E^{\rm b}_{353,\rm HM2}  =(46.6\pm1.1)\, \musqkcmb  \ , \label{eq:7.1}\\
V^{BB} \ ({\rm HL}) &=  \frac{1}{N_{\rm HL}} \sum_{i=1}^{N_{\rm HL}} B^{\rm b}_{353,\rm HM1} B^{\rm b}_{353,\rm HM2}  = (29.1\pm1.0)\, \musqkcmb \ , \label{eq:7.2}
\end{align}
where $N_{\rm HL}$ is the total number of pixels in the HL region. 
The ratio of the filtered \B\ and \E\ variances is 
\begin{equation}
\frac{V^{BB}\ ({\rm HL})}{V^{EE}\ ({\rm HL})}  =0.62 \pm 0.03 \ .
\end{equation}
The uncertainty on $V^{EE}$ is computed by repeating the calculation of Eq.~\eqref{eq:7.1} using the different cross-products, i.e., the two HalfRing (HR1 and HR2) and the two DetSet (DS1 and DS2) maps. We use the 
cross-HalfMissions as a reference for mean $V^{EE}$.  The $1\,\sigma$ uncertainty on $V^{EE}$ comes from the differences of these cross-products (DetSets minus HalfMissions and HalfRings minus HalfMissions). 
This $1\,\sigma$ uncertainty is dominated by data systematics rather than statistical noise. This is in agreement with the uncertainties on power spectra over the same $\ell$ range, as shown for $f_{\rm sky} =0.5$
in Figure 2 of \citet{planck2014-XXX}. The statistical noise on $V^{EE}$ is estimated from the cross-HalfMissions between their two HalfRing half-differences, 
$N^{EE}= (E_{353, \rm HR1}^{\rm b} - E_{353, \rm HR2}^{\rm b})_{\rm HM1}/2 \times  (E_{353, \rm HR1}^{\rm b} - E_{353, \rm HR2}^{\rm b})_{\rm HM2}/2$  as
\begin{equation}
\sigma_{V^{EE}} \ ({\rm HL})  = \frac{\sigma_{N^{EE}}}{\sqrt{N_{\rm HL}}} = 0.02 \, \musqkcmb \ .
\end{equation}
A similar procedure is applied to compute the uncertainty on $V^{BB}$. The ratio of the \B\ and \E\ variances differs slightly from the measurement at the power spectra level \citep{planck2014-XXX}, probably because of the multipole range over which the ratio of the \B\ and \E\ variances are computed.

We compute the covariances of the three maps over the high-latitude sky using the relations
\begin{align}
V^{TE} \ ({\rm HL}) &=  \frac{1}{N_{\rm HL}} \sum_{i=1}^{N_{\rm HL}} \fDmodel\ \fE\  =(124.1\pm 1.4)\, \musqkcmb  \ , \\
V^{TB} \ ({\rm HL}) &=  \frac{1}{N_{\rm HL}} \sum_{i=1}^{N_{\rm HL}} \fDmodel\ \fB\ =(3.0\pm1.5)\, \musqkcmb \ , \label{eq:7.4}\\
V^{EB} \ ({\rm HL}) &=  \frac{1}{N_{\rm HL}} \sum_{i=1}^{N_{\rm HL}} \fE\ \fB\ =(-0.2\pm0.3)\, \musqkcmb \ . \label{eq:7.5}
\end{align}
The above covariances divided by $V^{EE}$ are listed in Table~\ref{tab:7.1}. The ratio $V^{TE}/V^{EE}$ that we find is consistent with the measurements of \citet{planck2014-XXX} at the power spectrum level.

\subsection{Contribution of filaments to the variance of the $E$ and $B$ maps}\label{sec:6.3}

\begin{figure}
\includegraphics[width=8.8cm]{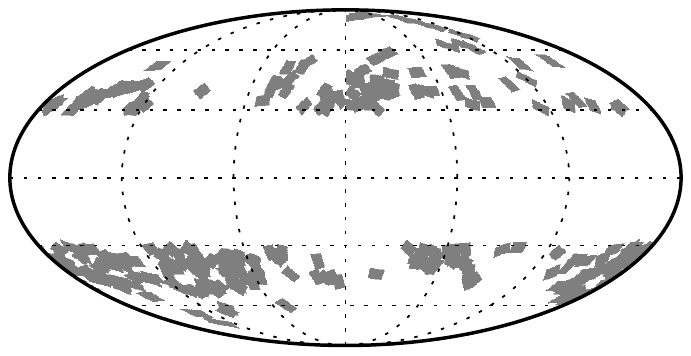}
\caption{Map of the selected pixels (grey colour) used in the stacking analysis. It covers 28\,\% of the high-latitude sky. Each tile in the image is a $7\deg \times 5\deg$ patch around the filament centre and rotated by \hessangle.}
\label{fig:6.3}
\end{figure}

In this section, we compute the variance of the \fE\ and \fB\ maps over the sky pixels used to produce the stacked images in Fig.~\ref{fig:6.1}, and compare the values with those measured over  the HL region. These pixels are within the $7\deg \times 5\deg$ patches, with an orientation angle \hessangle, centred on the 
filaments (Sect.~\ref{sec:5.1}). These pixels define the grey regions in Fig.~\ref{fig:6.3}. 
We label them as ${\rm SP}$ and the rest of the high-latitude sky as ${\rm O}$. The stacking procedure includes the filaments along with their surrounding background emission and, hence, effectively 
increases the selected fraction of the high-latitude sky, $f_1$, from 2.2\,\% (filament pixels as described in Sect.~\ref{sec:3.2}) to 28\,\%.

We compute the variance from the SP pixels using the relation given in Eq.~\eqref{eq:7.1},
\begin{equation}
V^{EE} \ (\rm {SP}) = (137.5\pm1.4) \, \musqkcmb  \ .
\end{equation}
The sky variance of the \fE\ map in the high-latitude sky can be written as the sum of contributions from SP ($f_1=0.28$) and O ($1-f_1=0.72$) regions. It is given by
\begin{equation}
V^{EE} \ ({\rm HL})  =  f_1\times V^{EE} \ ({\rm SP})  +  (1-f_1) \times V^{EE} \ ({\rm O}) \ .
\end{equation}
The ratio  ($R_{\rm {SP}}$) of the variance from the stacked pixels to the total sky variance is given by
\begin{equation}
R_{\rm {SP}}= \frac{f_{1} \times V^{EE} \  ({\rm SP})}{V^{EE} \ ({\rm HL}) }=  0.83 \ .
\end{equation}
The value of $R_{\rm {SP}}$ is expected to be high, since the filaments are bright structures on the sky. The pixels we used for stacking contribute 83\,\% of the total sky variance in the high-latitude sky. A similar result has been reported 
for the synchrotron emission, where bright filaments/shells are also measured to contribute most of the sky variance in polarization \citep{Vidal:2014}.

It has been noted that the structure in the \Planck\ $353$\,GHz dust polarization maps is not fully accounted for by the filaments  seen in the total dust intensity map. In particular, the local dispersion of the 
polarization angle shows structures in the polarization maps that have no counterpart in total intensity \citep{planck2014-XIX}. These structures are thought to trace morphology of \bpos\ uncorrelated with matter structures 
\citep{planck2014-XX}. However, as our $R_{\rm {SP}}$ value shows, these polarization structures do not contribute much to the variance of the dust polarization. 

In the same way as in Sect.~\ref{sec:6.2}, we compute the variances $V^{EE}$, $V^{BB}$, $V^{TE}$, $V^{TB}$, and $V^{EB}$ over the ${\rm SP}$ and ${\rm O}$ regions. Table~\ref{tab:7.1} presents the ratios of the 
variances computed over different sky regions. 

\begin{table}[h!]
\begingroup
\newdimen\tblskip \tblskip=5pt
\caption{\label{tab:7.1} The ratios of the variances computed from the selected pixels (SP) used in the stacking analysis and the rest (O) of the high-latitude sky (HL).}
\nointerlineskip
\vskip -3mm
\setbox\tablebox=\vbox{
   \newdimen\digitwidth 
   \setbox0=\hbox{\rm 0} 
   \digitwidth=\wd0 
   \catcode`*=\active 
   \def*{\kern\digitwidth}
   \newdimen\signwidth 
   \setbox0=\hbox{+} 
   \signwidth=\wd0 
   \catcode`!=\active 
   \def!{\kern\signwidth}
\halign{
\hbox to 0.4 in{#\leaderfil}\tabskip=3.5em&
\hfil #\hfil&
\hfil #\hfil&
\hfil #\hfil\tabskip=0pt\cr
\noalign{\doubleline \vskip 2pt}
\omit \hfil  Ratio \hfil& {\rm SP} & {\rm O}   & {\rm HL}  \cr
\noalign{\vskip 4pt\hrule\vskip 6pt}
${V^{BB}}/{V^{EE}}$ &   $!0.66 \pm 0.01$   & $0.51 \pm   0.05$ & $0.62  \pm      0.03$  \cr
\noalign{\vskip 4pt}
${V^{TE}}/{V^{EE}}$ &   $!2.74 \pm 0.04$   & $2.48 \pm   0.15$ & $2.67   \pm     0.07$  \cr
\noalign{\vskip 4pt}
${V^{TB}}/{V^{EE}}$ &  $-0.07 \pm 0.04$  & $0.12 \pm   0.04$  & $0.06   \pm     0.03$  \cr
\noalign{\vskip 4pt}
${V^{EB}}/{V^{EE}}$ &  $-0.01 \pm 0.02$  & $0.02  \pm  0.02$  & $0.00    \pm    0.01$  \cr
\noalign{\vskip 5pt\hrule\vskip 3pt}}}
\endPlancktablewide
\endgroup
\end{table}

\subsection{Analytical approximation}\label{sec:6.4}

Following  \citet{Zaldarriaga:2001} and  the description of the filaments in Eqs.~\eqref{eq:6.m1} and ~\eqref{eq:6.0}, 
we can express the $E$- and $B$-modes of a given filament as
\begin{align}
 \aE^{\rm F}&\simeq -\  \abFQ= \phantom{-} \ \bar{P}_{353}^{\rm F} \cos\ 2 \Delta_{\angloc - \hessangle}^{\rm F} \ ,  \\
 \aB^{\rm F}&\simeq -\ \abFU= -\  \bar{P}_{353}^{\rm F} \sin\ 2 \Delta_{\angloc - \hessangle}^{\rm F} \ ,
\end{align}
where $\bar{P}_{353}^{\rm F}$ is the mean polarization intensity of the filament. The direct relation between $ \aE^{\rm F}$ and  $\abFQ$, and $\aB^{\rm F}$ and $\abFU$ only holds for an idealised filament. For $N$ idealized filaments oriented arbitrarily on the sky with respect to the GN, the ratio between the variances of the $B$ and $E$ 
maps is given by 
\begin{align}
\frac{V^{BB}}{V^{EE}} = \frac{\langle \aB^{\rm F} \aB^{\rm F} \rangle}{\langle \aE^{\rm F} \aE^{\rm F} \rangle}  &= \frac{\sum \frac{2\ell+1}{4\pi} \ C_{\ell}^{BB} \ w_{\ell}^2}{\sum \frac{2\ell+1}{4\pi} \ C_{\ell}^{EE} \ w_{\ell}^2} \label{eq:7.3}\ ,
\end{align}
which is expanded in terms of power spectra $C_{\ell}^{BB}$ and $C_{\ell}^{EE}$ under the assumption of statistical isotropy and homogeneity. The beam window function ($w_{\ell}$) is the filter function. Both the observed 
$C_{\ell}^{BB}$ and $C_{\ell}^{EE}$ dust power spectra follow a power-law model with the same slope $\alpha$ \citep{planck2014-XXX}. This reduces Eq.~\eqref{eq:7.3} to
\begin{equation}
\frac{V^{BB}}{V^{EE}} = \frac{\sum (2\ell+1) \ A^{BB}\  \ell^{-\alpha} \ w_{\ell}^2}{\sum (2\ell+1) \ A^{EE} \ \ell^{-\alpha} \ w_{\ell}^2} =  \frac{A^{BB}}{A^{EE}}  \ .
\end{equation}

From the histogram of the upper panel of Fig.~\ref{fig:4.2}, the distribution of the angle $\Delta_{\angloc - \hessangle}^{\rm F}$ is known for our filament sample.  Similar to the assumption made in Sect.~\ref{sec:5.2}, we 
assume that all the filaments have the same polarized intensity and therefore
\begin{equation}
 \frac{A^{BB}}{A^{EE}}  \simeq  \frac{\langle \sin^2\ 2 \Delta_{\angloc - \hessangle}^{\rm F} \rangle}{\langle \cos^2\ 2 \Delta_{\angloc - \hessangle}^{\rm F} \rangle} = 0.66 \ .
\end{equation}
We have computed the ratio $A^{BB}/A^{EE}$ using the 
two independent subsets of the \planck\ data (HM maps) and find the same mean value of 0.66. This value of the $A^{BB}/A^{EE}$ ratio based on this analytical model matches the observed mean value of  $0.62\pm0.03$ (Sect.~\ref{sec:6.2}). We note that the model value  is directly inferred from 
the distribution of $\Delta_{\angloc - \hessangle}^{\rm F}$ for our filament sample. If the HRO of $\Delta_{\angloc - \hessangle}^{\rm F}$ was flat with uniform probability between $-90\deg$ and $+90\deg$, we would have found equal variances in both the \E\ and \B\  maps. 

In summary, we propose that the alignment between \bpos\ and the filament orientations accounts for the \EB\ asymmetry in the range of angular scales $30< \ell < 300$. The mean value of  $V^{BB}/V^{EE}$ for ${\rm O}$ region is consistent with the ${\rm HL}$ value within $1.9\,\sigma$ (Table~\ref{tab:7.1}). This shows that a similar alignment between the matter structures and \bpos\ can be inferred over the rest of the high-latitude sky. 
Some high Galactic latitude sky areas, such as the BICEP2 field \citep{BICEP2:2014}, do not include any of the strong filaments from our study. The Planck $353$\,GHz polarization maps do not have the required signal-to-noise ratio to measure the $A^{BB}/A^{EE}$  ratio for individual BICEP2-like fields \citep{planck2014-XXX}. Therefore more sensitive observations will be needed to test whether our interpretation is relevant there.

\section{Relation to Galactic astrophysics}\label{sec:gal_astrophysics}

In this section we place the paper results in the context of earlier studies about the filamentary structure of interstellar matter and its correlation with the Galactic magnetic field. 

Over the last decades, observations of interstellar gas and dust  have been revealing the filamentary structure of the interstellar medium  in increasing details.
Before \planck\ and \herschel, the discovery of the infrared cirrus with the IRAS and \hi\ all-sky surveys was a main milestone in our perception of the structure of the diffuse ISM \citep{Boulanger:1994,Kalberla:2009}.
At high Galactic latitude, the \planck\ dust emission is tightly correlated with \hi\ emission at local velocities \citep{planck2013-XVII}. In particular, all of the filaments in our sample have an
\hi\ counterpart. They are selected on the $353$\,GHz map but are seen at far-infrared wavelengths, in particular the IRAS 100\,\micron\  map.

The interstellar filaments seen in the \planck\ $353$\,GHz dust intensity map are not all straight. With our filament-finding algorithm, we have identified the straight 
segments with lengths $L \ge 2\deg$. Some of these segments are pieces of longer non-straight filaments. The 259 filaments in our sample make most of the sky variance 
in polarization as measured in our analysis. This is not a complete sample but other filaments at high Galactic latitude do not contribute much to the dust power in $E$-modes.

A number of  studies, starting with the pioneering work of \citet{Goodman:1990}, have used the polarization of background starlight  to investigate the relative orientation between 
the interstellar filaments and \bpos. While most studies have targeted filaments identified in extinction maps of molecular clouds in the solar neighbourhood, a few studies 
have focussed on filaments seen in \hi\ emission \citep{McClure:2006,Clark:2014}. These last two papers report a preferred  alignment between the filaments  and \bpos\ in the diffuse ISM.
The analysis of \planck\ data has complemented earlier studies providing greater statistics and sensitivity. \citet{planck2014-XXXII} compare the orientations of matter
structures identified in the \planck\ $353$\,GHz map with that of the Galactic magnetic field at intermediate latitudes. The alignment between the filaments and \bpos\ reported 
in this paper becomes weaker for increasing column density (see Figure 15 of \citealt{planck2014-XXXII}).  
Towards molecular clouds the relative orientation is observed to change progressively from preferentially parallel in areas with the lowest column density to preferentially perpendicular in the 
areas with the highest column density \citep{planck2015-XXXV}. The transition occurs at a column density of $\rm 10^{21.7}\,cm^{-2}$. All the filaments considered in our analysis are much below 
this transition limit and observed, as expected from earlier studies, to be statistically aligned with \bpos.

\citet{planck2014-XXXII} and \citet{planck2015-XXXV} discuss these observational results in light of MHD simulations, which quantify the respective roles of the magnetic field, turbulence, 
and gas self-gravity in the formation of structures in the magnetized ISM. The alignment between the magnetic field and matter structures in the diffuse ISM is thought to be a signature of turbulence. 
Simulations show that turbulent flows will tend to stretch gas condensations into sheets and filaments, which appear elongated in column density maps \citep{Hennebelle:2013}. 
These structures will tend to be aligned with the magnetic field where the gas velocity is dynamically aligned with the field \citep{Brandenburg:2013}. Alignment also results from the fact 
that matter and the magnetic field are stretched in the same direction because the field is frozen into matter. The change in relative orientation observed within molecular clouds might be a 
signature of the formation of gravitationally bound structures in the presence of a dynamically important magnetic field. Indeed,  \citet{Soler:2013} report a change in the relative orientation between 
matter structures and the magnetic field, from parallel to perpendicular, for gravitationally bound structures in MHD simulations. This change is most significant for their simulation with the highest 
magnetization.

\section{Conclusion}\label{sec:conclusion}

We present a statistical study of the filamentary structure of the $353$\,GHz \Planck\ Stokes maps at high Galactic latitude, relevant to the study of dust emission as a polarization foreground to the CMB. The main results of our work are summarized as follows. 

We filter the intensity and polarization maps to isolate structures over the range of angular scales where the \EB\ power asymmetry is observed. 
From a Hessian analysis of the \Planck\ total dust intensity map at $353\,$GHz, we identify a sample of 259 filaments in the high-latitude sky with lengths $L \ge 2\deg$. We measure the mean orientation angle of each filament in this sample and find that the filaments are statistically aligned with the plane of the sky component of the magnetic field, \bpos,  inferred from the polarization angles measured by \planck. We also find that the orientation of \bpos\ is 
correlated with that of \bmean\ in the solar neighbourhood. Our results show that the correlation between the structures of interstellar matter and \bpos\ in the diffuse ISM  reported in
\citet{planck2014-XXXII} for intermediate Galactic latitudes also applies to the lower column density  filaments (a few $\rm 10^{19}\,cm^{-2}$)  observed at high Galactic latitude.

We present mean images of our filament sample in dust intensity and Stokes \Q\ and \U\ with respect to the filament orientation (\bQ\  and \bU), 
computed by stacking individual $7\deg \times 5\deg$ patches centred on each filament.  The stacked images show that the contribution of the filaments is a negative feature with
respect to the background in the \bQ\ image and is not seen in the \bU\ image. This result directly follows from the fact that the histogram of relative orientation between the filaments and \bpos\ peaks and is symmetric around 0\deg. Combining the stacked images and the histogram, we estimate the mean polarization fraction of the filaments to be 11\,\%. 

We relate the \EB\ asymmetry discovered in the power spectrum analysis of \Planck\ 353\,GHz polarization maps \citep{planck2014-XXX} to the alignment between the filaments and \bpos\ in the diffuse 
ISM. The set of  $7\deg \times 5\deg$ patches we stack represents 28\,\% of the sky area at high Galactic latitude. The power of the $E$-mode dust polarization computed over this area amounts to 83\,\% of the total 
dust polarization power in the high-latitude sky. We show with an analytical approximation of the filaments (based on the work of \citealt{Zaldarriaga:2001}), that the HRO between the filaments and \bpos\ 
may account for the $C_{\ell}^{BB}/C_{\ell}^{EE}$ ratio measured over the high-latitude sky. Our interpretation could also apply to the  \EB\ asymmetry reported for the synchrotron emission \citep{planck2014-a12}, since there is also a correlation between the orientation angle of \bpos\ and the filamentary structures of the synchrotron intensity map \citep{Vidal:2014, planck2014-a31}. 

Present models of the dust polarization sky \citep[e.g.,][]{ODea:2012,PSM:2013} produce an equal amount of power in $E$- and $B$-modes for masks excluding the Galactic plane, because they ignore 
the correlation between  the structure of the magnetic field and that of matter. Our work should motivate a quantitative modelling of the polarized sky, which will take into account the observed correlations between 
the Galactic magnetic field and the structure of interstellar matter.

\begin{acknowledgements}

The Planck Collaboration acknowledges the support of: ESA; CNES, and
CNRS/INSU-IN2P3-INP (France); ASI, CNR, and INAF (Italy); NASA and DoE
(USA); STFC and UKSA (UK); CSIC, MINECO, JA and RES (Spain); Tekes, AoF,
and CSC (Finland); DLR and MPG (Germany); CSA (Canada); DTU Space
(Denmark); SER/SSO (Switzerland); RCN (Norway); SFI (Ireland);
FCT/MCTES (Portugal); ERC and PRACE (EU). A description of the Planck
Collaboration and a list of its members, indicating which technical
or scientific activities they have been involved in, can be found at \href{http://www.cosmos.esa.int/web/planck/planck-collaboration}{http://www.cosmos.esa.int/web/planck/planck-collaboration}. 
The research leading to these results has received funding from the European Research
Council under the European Union's Seventh Framework Programme (FP7/2007-2013) / ERC grant agreement n$^\circ$ 267934. 
Some of the results in this paper have been derived using the \healpix\ package.

\end{acknowledgements}

\bibliographystyle{aat}
\input{DustEB.bbl}


\appendix
\section{Hessian analysis} \label{sec:hessian}

In this appendix, we detail the implementation of the Hessian analysis to determine the orientations of the filaments. We start with the  \fDmodel\ map at \healpix\  resolution $\Nside=512$. For each pixel on the
 sky, we estimate the first and second derivatives of \fDmodel\ with respect to the Galactic longitude $l$ and latitude $b$ (as described in \citealt{Monteserin:2005}). The Hessian matrix of  \fDmodel\  is defined as
\begin{align}
\large{H} & = \large{\begin{bmatrix} H_{\rm xx} & H_{\rm xy}\\
            H_{\rm xy} & H_{\rm yy} \end{bmatrix}}, \label{eq:A.1.1}
\end{align} 
where
\begin{align}
H_{\rm xx}  & = \frac{\partial^2 \fDmodel}{\partial^2 b} \ , \\
H_{\rm xy} & = \frac{\partial^2 \fDmodel}{\cos b \  \partial b \ \partial l}  \ , \\
H_{\rm yy} & =  \frac{\partial^2 \fDmodel}{\cos^2 b \  \partial^2 l} \ .
\end{align}      

We decompose the  \fDmodel\ map into $a_{\ell m}$ coefficients using the ``\ianafast" routine of \healpix\ and then use the ``\isynfast" routine in \healpix\ 
to compute the second partial derivatives at each pixel. This method is computationally faster than the one used in \citet{planck2014-XXXII}. The smallest 
eigenvalue, \mlambda, of the Hessian matrix is calculated as
\begin{equation}
\mlambda=\frac{1}{2} (H_{\rm xx}+H_{\rm yy} - \alpha) \ , \label{eq:A.1.3}
\end{equation}
where $\alpha = \sqrt{(H_{\rm xx} - H_{\rm yy})^2 + 4 H_{\rm xy}^2}$. The map of \mlambda, displayed in the lower left panel of Fig.~\ref{fig:3.1}, highlights the filaments in the \fDmodel\ map. The filament orientation angle, \mtheta, is calculated using the relation
\begin{equation}
\mtheta=  \text{atan} \left [ -\frac{H_{\rm xx} -H_{\rm yy} + \alpha}{2H_{\rm xy}} \right] \ . \label{eq:A.1.5}
\end{equation}
This follows the IAU convention, since it is measured from the GN and positive to the East direction. The formula for the filament orientation angle \mtheta\ is equivalent to equation ($9$) of \citet{planck2014-XXXII}.

\begin{figure}[ht!]
\begin{tabular}{c}
\includegraphics[width=8.8cm]{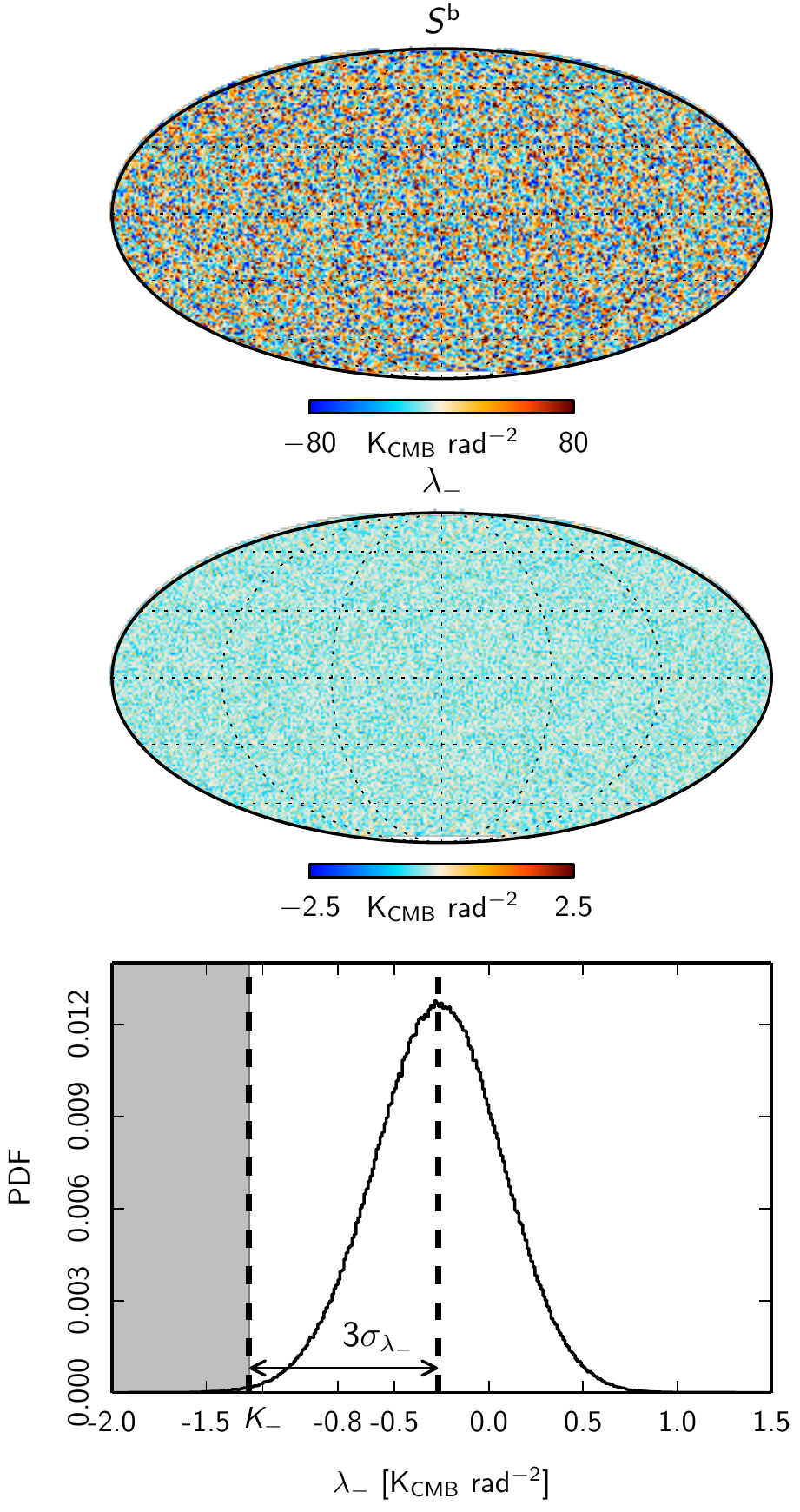}
\end{tabular}
\caption{Filtered realization of the Gaussian dust sky (upper panel) and the corresponding eigenvalue \mlambda\ (middle panel). The distribution of the eigenvalue \mlambda\ (lower panel) over the unmasked pixels is Gaussian.  }
\label{fig:B.1}
\end{figure}

\section{Gaussian realization of the dust sky} \label{sec:gauss_simul}

Here we apply the \smaff\ algorithm on a Gaussian realization of the dust total intensity map using the same methodology as described in Sect.~\ref{sec:3.2}. The power spectrum is modelled as 
$C_{\ell}^{TT} \propto \ell^{\alpha}$, where $\alpha$ is the slope of the power law.  We choose $\alpha=-2.4$ in the multipole range from $\ell=30$ to 300 \citep{planck2014-XXII}. 

We compute a Gaussian realization of a sky map on a \healpix\  grid with an $\ell^{-2.4}$ power spectrum, and use the spline wavelet transform to filter it. Hereafter, we refer to the filtered simulated Gaussian 
map as  the ``$S^{\rm b}$ map".  We compute the smallest eigenvalue \mlambda\ of the $S^{\rm b}$ map using the Hessian analysis described in Appendix~\ref{sec:hessian}. The $S^{\rm b}$ map and its corresponding \mlambda\ map are shown in the upper and middle panels of Fig.~\ref{fig:B.1}. 

We only consider the high-latitude sky. With the upper threshold $K_{-} = m_{\mlambda} - 3\,\sigma_{\lambda_{-}}$,  we remove pixels as shown in the lower panel of Fig.~\ref{fig:B.1}. Running our 
friend-of-friend algorithm on the $S_{\phantom{-}}^{\rm b}$ map, we do not detect any filament with $L \ge 2\deg$. For  higher values of the threshold $K_{-}$, we detect such filaments in the $S^{\rm b}$ map. 
We call them ``weak" filaments. The threshold factor $K_{-}$, used in our study, is a key factor to separate the strong filaments from the weak ones. The main results of this paper follow from the 
statistical properties of strong filaments. However, in Appendix~\ref{sec:vary_threshold} we demonstrate that they still hold for the weak filaments.

\section{Effect of \mlambda\ thresholding on the filament count} \label{sec:vary_threshold}

We apply the threshold $K_{-} = m_{\mlambda} - 3\,\sigma_{\mlambda}$ in Sect.~\ref{sec:3.2} to find strong filaments  in the \fDmodel\ map. In this section, we change the threshold $K_{-}$ to quantify its 
effect on the HRO between the filaments and \bpos. 
We choose different thresholds and divide the selected \mlambda\ coherent structures into two categories, namely strong and weak filaments. The strong filaments are  selected from the sky pixels that have $\mlambda < m_{\mlambda} - 3\,\sigma_{\mlambda}$. Whereas, the weak filaments are selected from the sky pixels that have $m_{\mlambda} - 3\,\sigma_{\mlambda} \le \mlambda < m_{\mlambda} - p\,\sigma_{\mlambda}$, where $p$ is a factor to defined the threshold $K_{-} = m_{\mlambda} - p\,\sigma_{\mlambda}$. In Table~\ref{tab:C.1}, we list the number of weak filaments for different values of $p$.

\begin{figure}[ht!]
\includegraphics[width=8.8cm]{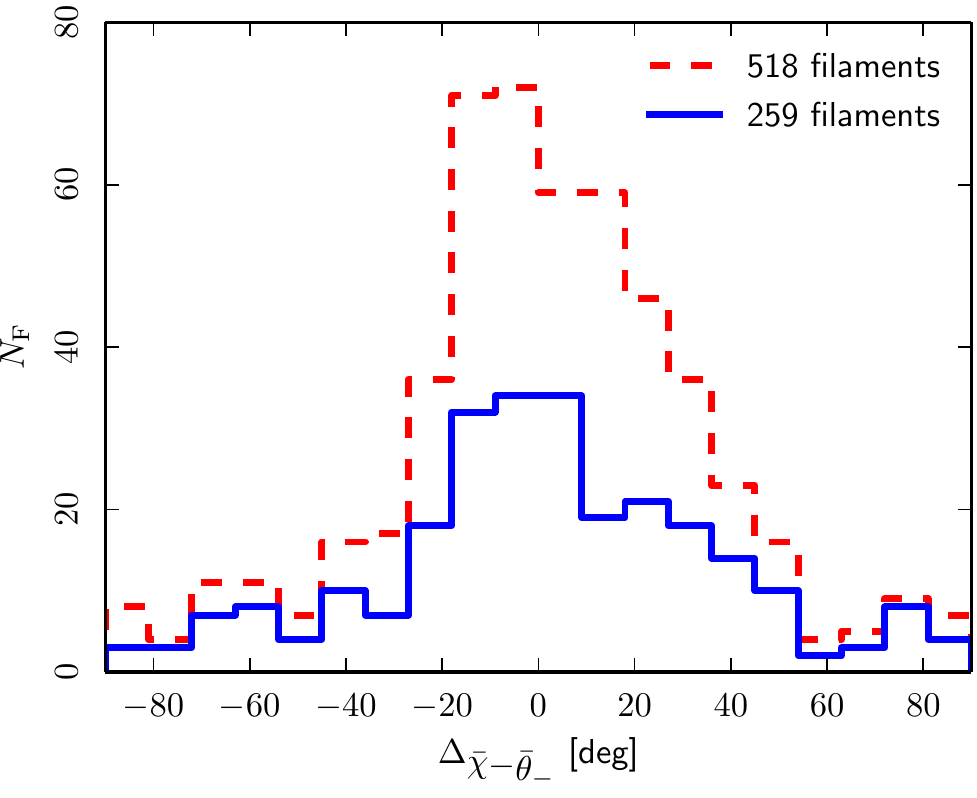}
\caption{Same as the upper panel of Fig.~\ref{fig:4.2}, when including angle differences $\Delta_{\angloc - \hessangle}$ from the weak filaments. The width of the histogram does not change.}
\label{fig:C.1}
\end{figure}

By  choosing a threshold $K_{-} = m_{\mlambda} - 1\,\sigma_{\mlambda}$ and running our friend-of-friend algorithm on the \fDmodel\ map, we double the number of filaments with $L\ge 2\deg$. For this larger set, we compute the HRO between the filaments and \bpos, and present it in Fig.~\ref{fig:C.1}. The $1\,\sigma$ dispersion of the angle difference remains the same as for our nominal set. This shows that the alignment between the structures of interstellar matter and \bpos\ holds even when including the weak filaments.

Including the weak filaments and their surrounding background emission in the analysis of Sect.~\ref{sec:6.3} effectively increases the sky fraction from 28\,\% to 50\,\% of the high-latitude sky. With more sky coverage, the ratio of the variances, i.e, $V^{BB}/V^{EE}$, $V^{TE}/V^{EE}$, $V^{TB}/V^{EE}$,  and $V^{EB}/V^{EE}$ is close to the HL region values quoted in Table~\ref{tab:7.1}.

\begin{table}[h!]
\begingroup
\newdimen\tblskip \tblskip=5pt
\caption{\label{tab:C.1} Total filament count, including the strong and weak filaments, as a function of the threshold $K_{-}$ on the \mlambda\ map derived from the \fDmodel\ map.}
\nointerlineskip
\vskip -3mm
\setbox\tablebox=\vbox{
   \newdimen\digitwidth 
   \setbox0=\hbox{\rm 0} 
   \digitwidth=\wd0 
   \catcode`*=\active 
   \def*{\kern\digitwidth}
   \newdimen\signwidth 
   \setbox0=\hbox{+} 
   \signwidth=\wd0 
   \catcode`!=\active 
   \def!{\kern\signwidth}
   \newdimen\pointwidth
   \setbox0=\hbox{{.}}
   \pointwidth=\wd0
   \catcode`?=\active
   \def?{\kern\pointwidth}
\halign{
\hbox to 0.2 in{#\leaderfil}\tabskip=0.4em&
\hfil #\hfil&
\hfil #\hfil&
\hfil #\hfil\tabskip=0pt\cr
\noalign{\doubleline \vskip 0pt}
\omit \hfil Factor \hfil& \multispan{3}\hfil Filament count \hfil\cr
\omit \hfil $p$ \hfil& weak & strong& total \cr
\omit \hfil \hfil& $m_{\mlambda} - 3\,\sigma_{\mlambda} \le \mlambda <  m_{\mlambda} - p\,\sigma_{\mlambda}$ &$\mlambda <  m_{\mlambda} - 3\,\sigma_{\mlambda}$ & \cr
\noalign{\vskip 4pt\hrule\vskip 6pt}
3 &     0    & 259 & 259  \cr
2 &   53    & 259 & 312  \cr
1.5 &  140  &  259 &399  \cr
1 &  259  &  259 &518  \cr
\noalign{\vskip 5pt\hrule\vskip 3pt}}}
\endPlancktablewide
\endgroup
\end{table}

\section{Uncertainties on the angles \hessangle\ and \angloc\ } \label{sec:angle_errors}

\begin{figure}[ht!]
\includegraphics[width=8.8cm]{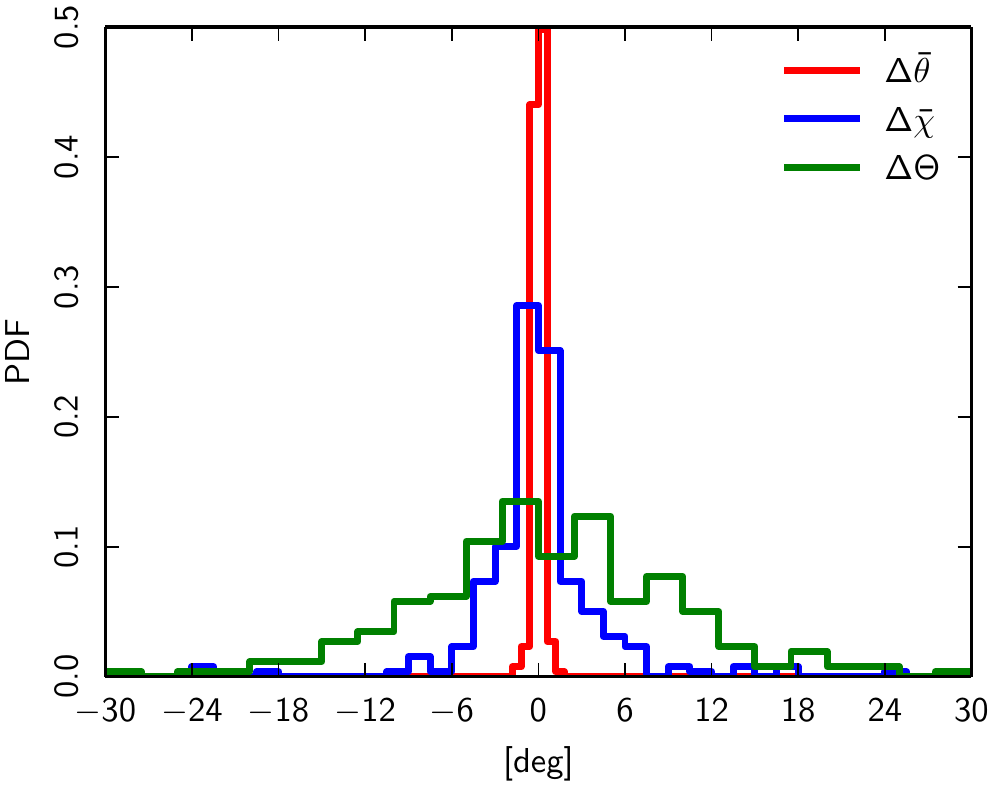}
\caption{Histogram of changes in \hessangle\ (red) and \angloc\ (blue) due to noise present in the \planck\ data.  The histogram of changes in \hessangle\ from the difference of the filtered and unfiltered \Dmodel\ maps is shown in green.}
\label{fig:D.1}
\end{figure}

In Sect.~\ref{sec:hro} we present our statistical analysis of  the relative orientation between the filaments and \bpos. The filament orientation angle, \hessangle, is a single number computed from the 
\fDmodel\ map. To estimate the error bar on \hessangle, we propagate the noise in the \Dmodel\ map ($N_{D_{\rm 353}}$)  through the  Hessian analysis.  We simulate a Gaussian realization of the $N_{D_{\rm 353}}$ map 
and add it to the \Dmodel\ map. We filter the noise added to the \Dmodel\ map in the same manner as the \Dmodel\ map. For each filament in our sample, we compute the change of its orientation angle from the added 
noise using the formula
\begin{equation}
\Delta \hessangle = \hessangle\ (\fDmodel + N_{D_{\rm 353}}^{\rm b})  -  \hessangle\ (\fDmodel).
\end{equation}
The distribution of the angle difference is presented with red colour in Fig.~\ref{fig:D.1}. The mean of the distribution is $0\deg$ and the $1\,\sigma$ dispersion is 0.2\deg. This shows that the noise in the \Dmodel\ map has little impact on our estimate of the filament orientation angle. 

Next, we compute the uncertainty on the orientation angle of \bpos. We use the two HM \planck\ polarization maps, and compute the orientation angle of \bpos\ using Eq.~\eqref{eq:2.2}. For each filament in our sample, we 
compute the difference between the two values obtained from the two HM maps,
\begin{equation}
\Delta \angloc = 0.5 \times [ \angloc_{\rm HM1}  -  \angloc_{\rm HM2} ].
\end{equation}
The histogram of the angle difference is presented with blue colour in Fig.~\ref{fig:D.1}. The mean of the distribution is $0\deg$ and the $1\,\sigma$ dispersion is 2\deg. The error on the orientation angle of \bpos\ is small compared to the width of the HROs in Fig.~\ref{fig:4.2}. 

Last, we assess the impact of the map filtering on our analysis, computing the mean orientations of the filaments,  over the same set of selected pixels, on the \Dmodel\ map. We compute the angle difference of the filament orientation derived from the filtered and unfiltered \planck\ 353 GHz maps 
\begin{equation}
\Delta \Theta = \hessangle\ (\Dmodel^{})  -  \hessangle\ (\fDmodel).
\end{equation}
The histogram of $\Delta \Theta$ is plotted with green colour in Fig.~\ref{fig:D.1}. The mean of the distribution is $0\deg$ and the $1\,\sigma$ dispersion is 7.0\deg. All the uncertainties measured in \hessangle\ and \angloc\ are small compared to the dispersion of the HRO measured in Fig.~\ref{fig:4.2}. This means that the data noise (for the intensity and polarization maps) and the filtering of the data are not critical for our study. 

\raggedright
\end{document}

%% file: Planck.tex
\def\setsymbol#1#2{\expandafter\def\csname #1\endcsname{#2}}
\def\getsymbol#1{\csname #1\endcsname}

\def\Planck{\textit{Planck}}

\def\all2013resultspapers{\nocite{planck2013-p01, planck2013-p02, planck2013-p02a, planck2013-p02d, planck2013-p02b, planck2013-p03, planck2013-p03c, planck2013-p03f, planck2013-p03d, planck2013-p03e, planck2013-p01a, planck2013-p06, planck2013-p03a, planck2013-pip88, planck2013-p08, planck2013-p11, planck2013-p12, planck2013-p13, planck2013-p14, planck2013-p15, planck2013-p05b, planck2013-p17, planck2013-p09, planck2013-p09a, planck2013-p20, planck2013-p19, planck2013-pipaberration, planck2013-p05, planck2013-p05a, planck2013-pip56, planck2013-p06b}}

\newbox\tablebox    \newdimen\tablewidth
\def\leaderfil{\leaders\hbox to 5pt{\hss.\hss}\hfil}
\def\endPlancktablewide{\tablewidth=\textwidth 
    $$\hss\copy\tablebox\hss$$
    \vskip-\lastskip\vskip -2pt}
\def\tablenote#1 #2\par{\begingroup \parindent=0.8em
    \abovedisplayshortskip=0pt\belowdisplayshortskip=0pt
    \noindent
    $$\hss\vbox{\hsize\tablewidth \hangindent=\parindent \hangafter=1 \noindent
    \hbox to \parindent{$^#1$\hss}\strut#2\strut\par}\hss$$
    \endgroup}
\def\doubleline{\vskip 3pt\hrule \vskip 1.5pt \hrule \vskip 5pt}

\def\L2{\ifmmode L_2\else $L_2$\fi}

\def\DeltaT{\ifmmode \Delta T\else $\Delta T$\fi}
\def\deltat{\ifmmode \Delta t\else $\Delta t$\fi}
\def\fknee{\ifmmode f_{\rm knee}\else $f_{\rm knee}$\fi}
\def\Fmax{\ifmmode F_{\rm max}\else $F_{\rm max}$\fi}
\def\solar{\ifmmode{\rm M}_{\mathord\odot}\else${\rm M}_{\mathord\odot}$\fi}
\def\Msolar{\ifmmode{\rm M}_{\mathord\odot}\else${\rm M}_{\mathord\odot}$\fi}
\def\Lsolar{\ifmmode{\rm L}_{\mathord\odot}\else${\rm L}_{\mathord\odot}$\fi}

\def\inv{\ifmmode^{-1}\else$^{-1}$\fi}
\def\mo{\ifmmode^{-1}\else$^{-1}$\fi}
\def\sup#1{\ifmmode ^{\rm #1}\else $^{\rm #1}$\fi}
\def\expo#1{\ifmmode \times 10^{#1}\else $\times 10^{#1}$\fi}
\def\,{\thinspace}
\def\lsim{\mathrel{\raise .4ex\hbox{\rlap{$<$}\lower 1.2ex\hbox{$\sim$}}}}
\def\gsim{\mathrel{\raise .4ex\hbox{\rlap{$>$}\lower 1.2ex\hbox{$\sim$}}}}

\def\simprop{\mathrel{\raise .4ex\hbox{\rlap{$\propto$}\lower 1.2ex\hbox{$\sim$}}}}
\def\deg{\ifmmode^\circ\else$^\circ$\fi}
\def\pdeg{\ifmmode $\setbox0=\hbox{$^{\circ}$}\rlap{\hskip.11\wd0 .}$^{\circ}
          \else \setbox0=\hbox{$^{\circ}$}\rlap{\hskip.11\wd0 .}$^{\circ}$\fi}
\def\arcs{\ifmmode {^{\scriptstyle\prime\prime}}
          \else $^{\scriptstyle\prime\prime}$\fi}
\def\arcm{\ifmmode {^{\scriptstyle\prime}}
          \else $^{\scriptstyle\prime}$\fi}
\newdimen\sa  \newdimen\sb
\def\parcs{\sa=.07em \sb=.03em
     \ifmmode \hbox{\rlap{.}}^{\scriptstyle\prime\kern -\sb\prime}\hbox{\kern -\sa}
     \else \rlap{.}$^{\scriptstyle\prime\kern -\sb\prime}$\kern -\sa\fi}
\def\parcm{\sa=.08em \sb=.03em
     \ifmmode \hbox{\rlap{.}\kern\sa}^{\scriptstyle\prime}\hbox{\kern-\sb}
     \else \rlap{.}\kern\sa$^{\scriptstyle\prime}$\kern-\sb\fi}
\def\ra[#1 #2 #3.#4]{#1\sup{h}#2\sup{m}#3\sup{s}\llap.#4}
\def\dec[#1 #2 #3.#4]{#1\deg#2\arcm#3\arcs\llap.#4}
\def\deco[#1 #2 #3]{#1\deg#2\arcm#3\arcs}
\def\rra[#1 #2]{#1\sup{h}#2\sup{m}}

\def\dots{\relax\ifmmode \ldots\else $\ldots$\fi}
\def\WHzsr{\ifmmode $W\,Hz\mo\,sr\mo$\else W\,Hz\mo\,sr\mo\fi}
\def\mHz{\ifmmode $\,mHz$\else \,mHz\fi}
\def\GHz{\ifmmode $\,GHz$\else \,GHz\fi}
\def\mKs{\ifmmode $\,mK\,s$^{1/2}\else \,mK\,s$^{1/2}$\fi}
\def\muKs{\ifmmode \,\mu$K\,s$^{1/2}\else \,$\mu$K\,s$^{1/2}$\fi}
\def\muKRJs{\ifmmode \,\mu$K$_{\rm RJ}$\,s$^{1/2}\else \,$\mu$K$_{\rm RJ}$\,s$^{1/2}$\fi}
\def\muKHz{\ifmmode \,\mu$K\,Hz$^{-1/2}\else \,$\mu$K\,Hz$^{-1/2}$\fi}
\def\MJysr{\ifmmode \,$MJy\,sr\mo$\else \,MJy\,sr\mo\fi}
\def\MJysrmK{\ifmmode \,$MJy\,sr\mo$\,mK$_{\rm CMB}\mo\else \,MJy\,sr\mo\,mK$_{\rm CMB}\mo$\fi}
\def\microns{\ifmmode \,\mu$m$\else \,$\mu$m\fi}
\def\micron{\microns}
\def\muK{\ifmmode \,\mu$K$\else \,$\mu$\hbox{K}\fi}
\def\microK{\ifmmode \,\mu$K$\else \,$\mu$\hbox{K}\fi}
\def\muW{\ifmmode \,\mu$W$\else \,$\mu$\hbox{W}\fi}
\def\kms{\ifmmode $\,km\,s$^{-1}\else \,km\,s$^{-1}$\fi}
\def\kmsMpc{\ifmmode $\,\kms\,Mpc\mo$\else \,\kms\,Mpc\mo\fi}
\setsymbol{LFI:center:frequency:70GHz:units}{70.3\,GHz}
\setsymbol{LFI:center:frequency:44GHz:units}{44.1\,GHz}
\setsymbol{LFI:center:frequency:30GHz:units}{28.5\,GHz}

\setsymbol{LFI:center:frequency:70GHz}{70.3}
\setsymbol{LFI:center:frequency:44GHz}{44.1}
\setsymbol{LFI:center:frequency:30GHz}{28.5}

\setsymbol{LFI:center:frequency:LFI18:Rad:M:units}{71.7\GHz}
\setsymbol{LFI:center:frequency:LFI19:Rad:M:units}{67.5\GHz}
\setsymbol{LFI:center:frequency:LFI20:Rad:M:units}{69.2\GHz}
\setsymbol{LFI:center:frequency:LFI21:Rad:M:units}{70.4\GHz}
\setsymbol{LFI:center:frequency:LFI22:Rad:M:units}{71.5\GHz}
\setsymbol{LFI:center:frequency:LFI23:Rad:M:units}{70.8\GHz}
\setsymbol{LFI:center:frequency:LFI24:Rad:M:units}{44.4\GHz}
\setsymbol{LFI:center:frequency:LFI25:Rad:M:units}{44.0\GHz}
\setsymbol{LFI:center:frequency:LFI26:Rad:M:units}{43.9\GHz}
\setsymbol{LFI:center:frequency:LFI27:Rad:M:units}{28.3\GHz}
\setsymbol{LFI:center:frequency:LFI28:Rad:M:units}{28.8\GHz}
\setsymbol{LFI:center:frequency:LFI18:Rad:S:units}{70.1\GHz}
\setsymbol{LFI:center:frequency:LFI19:Rad:S:units}{69.6\GHz}
\setsymbol{LFI:center:frequency:LFI20:Rad:S:units}{69.5\GHz}
\setsymbol{LFI:center:frequency:LFI21:Rad:S:units}{69.5\GHz}
\setsymbol{LFI:center:frequency:LFI22:Rad:S:units}{72.8\GHz}
\setsymbol{LFI:center:frequency:LFI23:Rad:S:units}{71.3\GHz}
\setsymbol{LFI:center:frequency:LFI24:Rad:S:units}{44.1\GHz}
\setsymbol{LFI:center:frequency:LFI25:Rad:S:units}{44.1\GHz}
\setsymbol{LFI:center:frequency:LFI26:Rad:S:units}{44.1\GHz}
\setsymbol{LFI:center:frequency:LFI27:Rad:S:units}{28.5\GHz}
\setsymbol{LFI:center:frequency:LFI28:Rad:S:units}{28.2\GHz}

\setsymbol{LFI:center:frequency:LFI18:Rad:M}{71.7}
\setsymbol{LFI:center:frequency:LFI19:Rad:M}{67.5}
\setsymbol{LFI:center:frequency:LFI20:Rad:M}{69.2}
\setsymbol{LFI:center:frequency:LFI21:Rad:M}{70.4}
\setsymbol{LFI:center:frequency:LFI22:Rad:M}{71.5}
\setsymbol{LFI:center:frequency:LFI23:Rad:M}{70.8}
\setsymbol{LFI:center:frequency:LFI24:Rad:M}{44.4}
\setsymbol{LFI:center:frequency:LFI25:Rad:M}{44.0}
\setsymbol{LFI:center:frequency:LFI26:Rad:M}{43.9}
\setsymbol{LFI:center:frequency:LFI27:Rad:M}{28.3}
\setsymbol{LFI:center:frequency:LFI28:Rad:M}{28.8}
\setsymbol{LFI:center:frequency:LFI18:Rad:S}{70.1}
\setsymbol{LFI:center:frequency:LFI19:Rad:S}{69.6}
\setsymbol{LFI:center:frequency:LFI20:Rad:S}{69.5}
\setsymbol{LFI:center:frequency:LFI21:Rad:S}{69.5}
\setsymbol{LFI:center:frequency:LFI22:Rad:S}{72.8}
\setsymbol{LFI:center:frequency:LFI23:Rad:S}{71.3}
\setsymbol{LFI:center:frequency:LFI24:Rad:S}{44.1}
\setsymbol{LFI:center:frequency:LFI25:Rad:S}{44.1}
\setsymbol{LFI:center:frequency:LFI26:Rad:S}{44.1}
\setsymbol{LFI:center:frequency:LFI27:Rad:S}{28.5}
\setsymbol{LFI:center:frequency:LFI28:Rad:S}{28.2}

\setsymbol{LFI:white:noise:sensitivity:70GHz:units}{134.7\muKs}
\setsymbol{LFI:white:noise:sensitivity:44GHz:units}{164.7\muKs}
\setsymbol{LFI:white:noise:sensitivity:30GHz:units}{143.4\muKs}

\setsymbol{LFI:white:noise:sensitivity:70GHz}{134.7}
\setsymbol{LFI:white:noise:sensitivity:44GHz}{164.7}
\setsymbol{LFI:white:noise:sensitivity:30GHz}{143.4}

\setsymbol{LFI:white:noise:sensitivity:LFI18:Rad:M:units}{512.0\muKs}
\setsymbol{LFI:white:noise:sensitivity:LFI19:Rad:M:units}{581.4\muKs}
\setsymbol{LFI:white:noise:sensitivity:LFI20:Rad:M:units}{590.8\muKs}
\setsymbol{LFI:white:noise:sensitivity:LFI21:Rad:M:units}{455.2\muKs}
\setsymbol{LFI:white:noise:sensitivity:LFI22:Rad:M:units}{492.0\muKs}
\setsymbol{LFI:white:noise:sensitivity:LFI23:Rad:M:units}{507.7\muKs}
\setsymbol{LFI:white:noise:sensitivity:LFI24:Rad:M:units}{462.2\muKs}
\setsymbol{LFI:white:noise:sensitivity:LFI25:Rad:M:units}{413.6\muKs}
\setsymbol{LFI:white:noise:sensitivity:LFI26:Rad:M:units}{478.6\muKs}
\setsymbol{LFI:white:noise:sensitivity:LFI27:Rad:M:units}{277.7\muKs}
\setsymbol{LFI:white:noise:sensitivity:LFI28:Rad:M:units}{312.3\muKs}
\setsymbol{LFI:white:noise:sensitivity:LFI18:Rad:S:units}{465.7\muKs}
\setsymbol{LFI:white:noise:sensitivity:LFI19:Rad:S:units}{555.6\muKs}
\setsymbol{LFI:white:noise:sensitivity:LFI20:Rad:S:units}{623.2\muKs}
\setsymbol{LFI:white:noise:sensitivity:LFI21:Rad:S:units}{564.1\muKs}
\setsymbol{LFI:white:noise:sensitivity:LFI22:Rad:S:units}{534.4\muKs}
\setsymbol{LFI:white:noise:sensitivity:LFI23:Rad:S:units}{542.4\muKs}
\setsymbol{LFI:white:noise:sensitivity:LFI24:Rad:S:units}{399.2\muKs}
\setsymbol{LFI:white:noise:sensitivity:LFI25:Rad:S:units}{392.6\muKs}
\setsymbol{LFI:white:noise:sensitivity:LFI26:Rad:S:units}{418.6\muKs}
\setsymbol{LFI:white:noise:sensitivity:LFI27:Rad:S:units}{302.9\muKs}
\setsymbol{LFI:white:noise:sensitivity:LFI28:Rad:S:units}{285.3\muKs}

\setsymbol{LFI:white:noise:sensitivity:LFI18:Rad:M}{512.0}
\setsymbol{LFI:white:noise:sensitivity:LFI19:Rad:M}{581.4}
\setsymbol{LFI:white:noise:sensitivity:LFI20:Rad:M}{590.8}
\setsymbol{LFI:white:noise:sensitivity:LFI21:Rad:M}{455.2}
\setsymbol{LFI:white:noise:sensitivity:LFI22:Rad:M}{492.0}
\setsymbol{LFI:white:noise:sensitivity:LFI23:Rad:M}{507.7}
\setsymbol{LFI:white:noise:sensitivity:LFI24:Rad:M}{462.2}
\setsymbol{LFI:white:noise:sensitivity:LFI25:Rad:M}{413.6}
\setsymbol{LFI:white:noise:sensitivity:LFI26:Rad:M}{478.6}
\setsymbol{LFI:white:noise:sensitivity:LFI27:Rad:M}{277.7}
\setsymbol{LFI:white:noise:sensitivity:LFI28:Rad:M}{312.3}
\setsymbol{LFI:white:noise:sensitivity:LFI18:Rad:S}{465.7}
\setsymbol{LFI:white:noise:sensitivity:LFI19:Rad:S}{555.6}
\setsymbol{LFI:white:noise:sensitivity:LFI20:Rad:S}{623.2}
\setsymbol{LFI:white:noise:sensitivity:LFI21:Rad:S}{564.1}
\setsymbol{LFI:white:noise:sensitivity:LFI22:Rad:S}{534.4}
\setsymbol{LFI:white:noise:sensitivity:LFI23:Rad:S}{542.4}
\setsymbol{LFI:white:noise:sensitivity:LFI24:Rad:S}{399.2}
\setsymbol{LFI:white:noise:sensitivity:LFI25:Rad:S}{392.6}
\setsymbol{LFI:white:noise:sensitivity:LFI26:Rad:S}{418.6}
\setsymbol{LFI:white:noise:sensitivity:LFI27:Rad:S}{302.9}
\setsymbol{LFI:white:noise:sensitivity:LFI28:Rad:S}{285.3}

\setsymbol{LFI:knee:frequency:70GHz:units}{29.5\mHz}
\setsymbol{LFI:knee:frequency:44GHz:units}{56.2\mHz}
\setsymbol{LFI:knee:frequency:30GHz:units}{113.7\mHz}

\setsymbol{LFI:knee:frequency:70GHz}{29.5}
\setsymbol{LFI:knee:frequency:44GHz}{56.2}
\setsymbol{LFI:knee:frequency:30GHz}{113.7}

\setsymbol{LFI:knee:frequency:LFI18:Rad:M:units}{16.3\mHz}
\setsymbol{LFI:knee:frequency:LFI19:Rad:M:units}{15.1\mHz}
\setsymbol{LFI:knee:frequency:LFI20:Rad:M:units}{18.7\mHz}
\setsymbol{LFI:knee:frequency:LFI21:Rad:M:units}{37.2\mHz}
\setsymbol{LFI:knee:frequency:LFI22:Rad:M:units}{12.7\mHz}
\setsymbol{LFI:knee:frequency:LFI23:Rad:M:units}{34.6\mHz}
\setsymbol{LFI:knee:frequency:LFI24:Rad:M:units}{46.2\mHz}
\setsymbol{LFI:knee:frequency:LFI25:Rad:M:units}{24.9\mHz}
\setsymbol{LFI:knee:frequency:LFI26:Rad:M:units}{67.6\mHz}
\setsymbol{LFI:knee:frequency:LFI27:Rad:M:units}{187.4\mHz}
\setsymbol{LFI:knee:frequency:LFI28:Rad:M:units}{122.2\mHz}
\setsymbol{LFI:knee:frequency:LFI18:Rad:S:units}{17.7\mHz}
\setsymbol{LFI:knee:frequency:LFI19:Rad:S:units}{22.0\mHz}
\setsymbol{LFI:knee:frequency:LFI20:Rad:S:units}{8.7\mHz}
\setsymbol{LFI:knee:frequency:LFI21:Rad:S:units}{25.9\mHz}
\setsymbol{LFI:knee:frequency:LFI22:Rad:S:units}{15.8\mHz}
\setsymbol{LFI:knee:frequency:LFI23:Rad:S:units}{129.8\mHz}
\setsymbol{LFI:knee:frequency:LFI24:Rad:S:units}{100.9\mHz}
\setsymbol{LFI:knee:frequency:LFI25:Rad:S:units}{38.9\mHz}
\setsymbol{LFI:knee:frequency:LFI26:Rad:S:units}{58.9\mHz}
\setsymbol{LFI:knee:frequency:LFI27:Rad:S:units}{104.4\mHz}
\setsymbol{LFI:knee:frequency:LFI28:Rad:S:units}{40.7\mHz}

\setsymbol{LFI:knee:frequency:LFI18:Rad:M}{16.3}
\setsymbol{LFI:knee:frequency:LFI19:Rad:M}{15.1}
\setsymbol{LFI:knee:frequency:LFI20:Rad:M}{18.7}
\setsymbol{LFI:knee:frequency:LFI21:Rad:M}{37.2}
\setsymbol{LFI:knee:frequency:LFI22:Rad:M}{12.7}
\setsymbol{LFI:knee:frequency:LFI23:Rad:M}{34.6}
\setsymbol{LFI:knee:frequency:LFI24:Rad:M}{46.2}
\setsymbol{LFI:knee:frequency:LFI25:Rad:M}{24.9}
\setsymbol{LFI:knee:frequency:LFI26:Rad:M}{67.6}
\setsymbol{LFI:knee:frequency:LFI27:Rad:M}{187.4}
\setsymbol{LFI:knee:frequency:LFI28:Rad:M}{122.2}
\setsymbol{LFI:knee:frequency:LFI18:Rad:S}{17.7}
\setsymbol{LFI:knee:frequency:LFI19:Rad:S}{22.0}
\setsymbol{LFI:knee:frequency:LFI20:Rad:S}{8.7}
\setsymbol{LFI:knee:frequency:LFI21:Rad:S}{25.9}
\setsymbol{LFI:knee:frequency:LFI22:Rad:S}{15.8}
\setsymbol{LFI:knee:frequency:LFI23:Rad:S}{129.8}
\setsymbol{LFI:knee:frequency:LFI24:Rad:S}{100.9}
\setsymbol{LFI:knee:frequency:LFI25:Rad:S}{38.9}
\setsymbol{LFI:knee:frequency:LFI26:Rad:S}{58.9}
\setsymbol{LFI:knee:frequency:LFI27:Rad:S}{104.4}
\setsymbol{LFI:knee:frequency:LFI28:Rad:S}{40.7}

\setsymbol{LFI:slope:70GHz:units}{$-1.03$\mHz}
\setsymbol{LFI:slope:44GHz:units}{$-0.89$\mHz}
\setsymbol{LFI:slope:30GHz:units}{$-0.87$\mHz}

\setsymbol{LFI:slope:70GHz}{$-1.03$}
\setsymbol{LFI:slope:44GHz}{$-0.89$}
\setsymbol{LFI:slope:30GHz}{$-0.87$}

\setsymbol{LFI:slope:LFI18:Rad:M:units}{$-1.04$\mHz}
\setsymbol{LFI:slope:LFI19:Rad:M:units}{$-1.09$\mHz}
\setsymbol{LFI:slope:LFI20:Rad:M:units}{$-0.69$\mHz}
\setsymbol{LFI:slope:LFI21:Rad:M:units}{$-1.56$\mHz}
\setsymbol{LFI:slope:LFI22:Rad:M:units}{$-1.01$\mHz}
\setsymbol{LFI:slope:LFI23:Rad:M:units}{$-0.96$\mHz}
\setsymbol{LFI:slope:LFI24:Rad:M:units}{$-0.83$\mHz}
\setsymbol{LFI:slope:LFI25:Rad:M:units}{$-0.91$\mHz}
\setsymbol{LFI:slope:LFI26:Rad:M:units}{$-0.95$\mHz}
\setsymbol{LFI:slope:LFI27:Rad:M:units}{$-0.87$\mHz}
\setsymbol{LFI:slope:LFI28:Rad:M:units}{$-0.88$\mHz}
\setsymbol{LFI:slope:LFI18:Rad:S:units}{$-1.15$\mHz}
\setsymbol{LFI:slope:LFI19:Rad:S:units}{$-1.00$\mHz}
\setsymbol{LFI:slope:LFI20:Rad:S:units}{$-0.95$\mHz}
\setsymbol{LFI:slope:LFI21:Rad:S:units}{$-0.92$\mHz}
\setsymbol{LFI:slope:LFI22:Rad:S:units}{$-1.01$\mHz}
\setsymbol{LFI:slope:LFI23:Rad:S:units}{$-0.95$\mHz}
\setsymbol{LFI:slope:LFI24:Rad:S:units}{$-0.73$\mHz}
\setsymbol{LFI:slope:LFI25:Rad:S:units}{$-1.16$\mHz}
\setsymbol{LFI:slope:LFI26:Rad:S:units}{$-0.79$\mHz}
\setsymbol{LFI:slope:LFI27:Rad:S:units}{$-0.82$\mHz}
\setsymbol{LFI:slope:LFI28:Rad:S:units}{$-0.91$\mHz}

\setsymbol{LFI:slope:LFI18:Rad:M}{$-1.04$}
\setsymbol{LFI:slope:LFI19:Rad:M}{$-1.09$}
\setsymbol{LFI:slope:LFI20:Rad:M}{$-0.69$}
\setsymbol{LFI:slope:LFI21:Rad:M}{$-1.56$}
\setsymbol{LFI:slope:LFI22:Rad:M}{$-1.01$}
\setsymbol{LFI:slope:LFI23:Rad:M}{$-0.96$}
\setsymbol{LFI:slope:LFI24:Rad:M}{$-0.83$}
\setsymbol{LFI:slope:LFI25:Rad:M}{$-0.91$}
\setsymbol{LFI:slope:LFI26:Rad:M}{$-0.95$}
\setsymbol{LFI:slope:LFI27:Rad:M}{$-0.87$}
\setsymbol{LFI:slope:LFI28:Rad:M}{$-0.88$}
\setsymbol{LFI:slope:LFI18:Rad:S}{$-1.15$}
\setsymbol{LFI:slope:LFI19:Rad:S}{$-1.00$}
\setsymbol{LFI:slope:LFI20:Rad:S}{$-0.95$}
\setsymbol{LFI:slope:LFI21:Rad:S}{$-0.92$}
\setsymbol{LFI:slope:LFI22:Rad:S}{$-1.01$}
\setsymbol{LFI:slope:LFI23:Rad:S}{$-0.95$}
\setsymbol{LFI:slope:LFI24:Rad:S}{$-0.73$}
\setsymbol{LFI:slope:LFI25:Rad:S}{$-1.16$}
\setsymbol{LFI:slope:LFI26:Rad:S}{$-0.79$}
\setsymbol{LFI:slope:LFI27:Rad:S}{$-0.82$}
\setsymbol{LFI:slope:LFI28:Rad:S}{$-0.91$}

\setsymbol{LFI:FWHM:70GHz:units}{13\parcm01}
\setsymbol{LFI:FWHM:44GHz:units}{27\parcm92}
\setsymbol{LFI:FWHM:30GHz:units}{32\parcm65}

\setsymbol{LFI:FWHM:70GHz}{13.01}
\setsymbol{LFI:FWHM:44GHz}{27.92}
\setsymbol{LFI:FWHM:30GHz}{32.65}

\setsymbol{LFI:FWHM:LFI18:units}{13\parcm39}
\setsymbol{LFI:FWHM:LFI19:units}{13\parcm01}
\setsymbol{LFI:FWHM:LFI20:units}{12\parcm75}
\setsymbol{LFI:FWHM:LFI21:units}{12\parcm74}
\setsymbol{LFI:FWHM:LFI22:units}{12\parcm87}
\setsymbol{LFI:FWHM:LFI23:units}{13\parcm27}
\setsymbol{LFI:FWHM:LFI24:units}{22\parcm98}
\setsymbol{LFI:FWHM:LFI25:units}{30\parcm46}
\setsymbol{LFI:FWHM:LFI26:units}{30\parcm31}
\setsymbol{LFI:FWHM:LFI27:units}{32\parcm65}
\setsymbol{LFI:FWHM:LFI28:units}{32\parcm66}

\setsymbol{LFI:FWHM:LFI18}{13.39}
\setsymbol{LFI:FWHM:LFI19}{13.01}
\setsymbol{LFI:FWHM:LFI20}{12.75}
\setsymbol{LFI:FWHM:LFI21}{12.74}
\setsymbol{LFI:FWHM:LFI22}{12.87}
\setsymbol{LFI:FWHM:LFI23}{13.27}
\setsymbol{LFI:FWHM:LFI24}{22.98}
\setsymbol{LFI:FWHM:LFI25}{30.46}
\setsymbol{LFI:FWHM:LFI26}{30.31}
\setsymbol{LFI:FWHM:LFI27}{32.65}
\setsymbol{LFI:FWHM:LFI28}{32.66}

\setsymbol{LFI:FWHM:uncertainty:LFI18:units}{0.170\arcm}
\setsymbol{LFI:FWHM:uncertainty:LFI19:units}{0.174\arcm}
\setsymbol{LFI:FWHM:uncertainty:LFI20:units}{0.170\arcm}
\setsymbol{LFI:FWHM:uncertainty:LFI21:units}{0.156\arcm}
\setsymbol{LFI:FWHM:uncertainty:LFI22:units}{0.164\arcm}
\setsymbol{LFI:FWHM:uncertainty:LFI23:units}{0.171\arcm}
\setsymbol{LFI:FWHM:uncertainty:LFI24:units}{0.652\arcm}
\setsymbol{LFI:FWHM:uncertainty:LFI25:units}{1.075\arcm}
\setsymbol{LFI:FWHM:uncertainty:LFI26:units}{1.131\arcm}
\setsymbol{LFI:FWHM:uncertainty:LFI27:units}{1.266\arcm}
\setsymbol{LFI:FWHM:uncertainty:LFI28:units}{1.287\arcm}

\setsymbol{LFI:FWHM:uncertainty:LFI18}{0.170}
\setsymbol{LFI:FWHM:uncertainty:LFI19}{0.174}
\setsymbol{LFI:FWHM:uncertainty:LFI20}{0.170}
\setsymbol{LFI:FWHM:uncertainty:LFI21}{0.156}
\setsymbol{LFI:FWHM:uncertainty:LFI22}{0.164}
\setsymbol{LFI:FWHM:uncertainty:LFI23}{0.171}
\setsymbol{LFI:FWHM:uncertainty:LFI24}{0.652}
\setsymbol{LFI:FWHM:uncertainty:LFI25}{1.075}
\setsymbol{LFI:FWHM:uncertainty:LFI26}{1.131}
\setsymbol{LFI:FWHM:uncertainty:LFI27}{1.266}
\setsymbol{LFI:FWHM:uncertainty:LFI28}{1.287}

\setsymbol{HFI:center:frequency:100GHz:units}{100\,GHz}
\setsymbol{HFI:center:frequency:143GHz:units}{143\,GHz}
\setsymbol{HFI:center:frequency:217GHz:units}{217\,GHz}
\setsymbol{HFI:center:frequency:353GHz:units}{353\,GHz}
\setsymbol{HFI:center:frequency:545GHz:units}{545\,GHz}
\setsymbol{HFI:center:frequency:857GHz:units}{857\,GHz}

\setsymbol{HFI:center:frequency:100GHz}{100}
\setsymbol{HFI:center:frequency:143GHz}{143}
\setsymbol{HFI:center:frequency:217GHz}{217}
\setsymbol{HFI:center:frequency:353GHz}{353}
\setsymbol{HFI:center:frequency:545GHz}{545}
\setsymbol{HFI:center:frequency:857GHz}{857}

\setsymbol{HFI:Ndetectors:100GHz}{8}
\setsymbol{HFI:Ndetectors:143GHz}{11}
\setsymbol{HFI:Ndetectors:217GHz}{12}
\setsymbol{HFI:Ndetectors:353GHz}{12}
\setsymbol{HFI:Ndetectors:545GHz}{3}
\setsymbol{HFI:Ndetectors:857GHz}{4}

\setsymbol{HFI:FWHM:Maps:100GHz:units}{9\parcm88}
\setsymbol{HFI:FWHM:Maps:143GHz:units}{7\parcm18}
\setsymbol{HFI:FWHM:Maps:217GHz:units}{4\parcm87}
\setsymbol{HFI:FWHM:Maps:353GHz:units}{4\parcm65}
\setsymbol{HFI:FWHM:Maps:545GHz:units}{4\parcm72}
\setsymbol{HFI:FWHM:Maps:857GHz:units}{4\parcm39}
\setsymbol{HFI:FWHM:Maps:100GHz}{9.88}
\setsymbol{HFI:FWHM:Maps:143GHz}{7.18}
\setsymbol{HFI:FWHM:Maps:217GHz}{4.87}
\setsymbol{HFI:FWHM:Maps:353GHz}{4.65}
\setsymbol{HFI:FWHM:Maps:545GHz}{4.72}
\setsymbol{HFI:FWHM:Maps:857GHz}{4.39}

\setsymbol{HFI:beam:ellipticity:Maps:100GHz}{1.15}
\setsymbol{HFI:beam:ellipticity:Maps:143GHz}{1.01}
\setsymbol{HFI:beam:ellipticity:Maps:217GHz}{1.06}
\setsymbol{HFI:beam:ellipticity:Maps:353GHz}{1.05}
\setsymbol{HFI:beam:ellipticity:Maps:545GHz}{1.14}
\setsymbol{HFI:beam:ellipticity:Maps:857GHz}{1.19}

\setsymbol{HFI:FWHM:Mars:100GHz:units}{9\parcm37}
\setsymbol{HFI:FWHM:Mars:143GHz:units}{7\parcm04}
\setsymbol{HFI:FWHM:Mars:217GHz:units}{4\parcm68}
\setsymbol{HFI:FWHM:Mars:353GHz:units}{4\parcm43}
\setsymbol{HFI:FWHM:Mars:545GHz:units}{3\parcm80}
\setsymbol{HFI:FWHM:Mars:857GHz:units}{3\parcm67}

\setsymbol{HFI:FWHM:Mars:100GHz}{9.37}
\setsymbol{HFI:FWHM:Mars:143GHz}{7.04}
\setsymbol{HFI:FWHM:Mars:217GHz}{4.68}
\setsymbol{HFI:FWHM:Mars:353GHz}{4.43}
\setsymbol{HFI:FWHM:Mars:545GHz}{3.80}
\setsymbol{HFI:FWHM:Mars:857GHz}{3.67}

\setsymbol{HFI:beam:ellipticity:Mars:100GHz}{1.18}
\setsymbol{HFI:beam:ellipticity:Mars:143GHz}{1.03}
\setsymbol{HFI:beam:ellipticity:Mars:217GHz}{1.14}
\setsymbol{HFI:beam:ellipticity:Mars:353GHz}{1.09}
\setsymbol{HFI:beam:ellipticity:Mars:545GHz}{1.25}
\setsymbol{HFI:beam:ellipticity:Mars:857GHz}{1.03}

\setsymbol{HFI:CMB:relative:calibration:100GHz}{$\lsim 1\%$}
\setsymbol{HFI:CMB:relative:calibration:143GHz}{$\lsim 1\%$}
\setsymbol{HFI:CMB:relative:calibration:217GHz}{$\lsim 1\%$}
\setsymbol{HFI:CMB:relative:calibration:353GHz}{$\lsim 1\%$}
\setsymbol{HFI:CMB:relative:calibration:545GHz}{}
\setsymbol{HFI:CMB:relative:calibration:857GHz}{}

\setsymbol{HFI:CMB:absolute:calibration:100GHz}{$\lsim 2\%$}
\setsymbol{HFI:CMB:absolute:calibration:143GHz}{$\lsim 2\%$}
\setsymbol{HFI:CMB:absolute:calibration:217GHz}{$\lsim 2\%$}
\setsymbol{HFI:CMB:absolute:calibration:353GHz}{$\lsim 2\%$}
\setsymbol{HFI:CMB:absolute:calibration:545GHz}{}
\setsymbol{HFI:CMB:absolute:calibration:857GHz}{}

\setsymbol{HFI:FIRAS:gain:calibration:accuracy:statistical:100GHz}{}
\setsymbol{HFI:FIRAS:gain:calibration:accuracy:statistical:143GHz}{}
\setsymbol{HFI:FIRAS:gain:calibration:accuracy:statistical:217GHz}{}
\setsymbol{HFI:FIRAS:gain:calibration:accuracy:statistical:353GHz}{2.5\%}
\setsymbol{HFI:FIRAS:gain:calibration:accuracy:statistical:545GHz}{1\%}
\setsymbol{HFI:FIRAS:gain:calibration:accuracy:statistical:857GHz}{0.5\%}

\setsymbol{HFI:FIRAS:gain:calibration:accuracy:systematic:100GHz}{}
\setsymbol{HFI:FIRAS:gain:calibration:accuracy:systematic:143GHz}{}
\setsymbol{HFI:FIRAS:gain:calibration:accuracy:systematic:217GHz}{}
\setsymbol{HFI:FIRAS:gain:calibration:accuracy:systematic:353GHz}{}
\setsymbol{HFI:FIRAS:gain:calibration:accuracy:systematic:545GHz}{7\%}
\setsymbol{HFI:FIRAS:gain:calibration:accuracy:systematic:857GHz}{7\%}

\setsymbol{HFI:FIRAS:zero:point:accuracy:100GHz:units}{0.8\MJysr}
\setsymbol{HFI:FIRAS:zero:point:accuracy:143GHz:units}{}
\setsymbol{HFI:FIRAS:zero:point:accuracy:217GHz:units}{}
\setsymbol{HFI:FIRAS:zero:point:accuracy:353GHz:units}{1.4\MJysr}
\setsymbol{HFI:FIRAS:zero:point:accuracy:545GHz:units}{2.2\MJysr}
\setsymbol{HFI:FIRAS:zero:point:accuracy:857GHz:units}{1.7\MJysr}

\setsymbol{HFI:FIRAS:zero:point:accuracy:100GHz}{0.8}
\setsymbol{HFI:FIRAS:zero:point:accuracy:143GHz}{}
\setsymbol{HFI:FIRAS:zero:point:accuracy:217GHz}{}
\setsymbol{HFI:FIRAS:zero:point:accuracy:353GHz}{1.4}
\setsymbol{HFI:FIRAS:zero:point:accuracy:545GHz}{2.2}
\setsymbol{HFI:FIRAS:zero:point:accuracy:857GHz}{1.7}

\setsymbol{HFI:unit:conversion:100GHz:units}{0.2415\MJysrmK}
\setsymbol{HFI:unit:conversion:143GHz:units}{0.3694\MJysrmK}
\setsymbol{HFI:unit:conversion:217GHz:units}{0.4811\MJysrmK}
\setsymbol{HFI:unit:conversion:353GHz:units}{0.2883\MJysrmK}
\setsymbol{HFI:unit:conversion:545GHz:units}{0.05826\MJysrmK}
\setsymbol{HFI:unit:conversion:857GHz:units}{0.002238\MJysrmK}

\setsymbol{HFI:unit:conversion:100GHz}{0.2415}
\setsymbol{HFI:unit:conversion:143GHz}{0.3694}
\setsymbol{HFI:unit:conversion:217GHz}{0.4811}
\setsymbol{HFI:unit:conversion:353GHz}{0.2883}
\setsymbol{HFI:unit:conversion:545GHz}{0.05826}
\setsymbol{HFI:unit:conversion:857GHz}{0.002238}

\setsymbol{HFI:colour:correction:alpha=-2:V1.01:100GHz}{0.9893}
\setsymbol{HFI:colour:correction:alpha=-2:V1.01:143GHz}{0.9759}
\setsymbol{HFI:colour:correction:alpha=-2:V1.01:217GHz}{1.0007}
\setsymbol{HFI:colour:correction:alpha=-2:V1.01:353GHz}{1.0028}
\setsymbol{HFI:colour:correction:alpha=-2:V1.01:545GHz}{1.0019}
\setsymbol{HFI:colour:correction:alpha=-2:V1.01:857GHz}{0.9889}

\setsymbol{HFI:colour:correction:alpha=0:V1.01:100GHz}{1.0008}
\setsymbol{HFI:colour:correction:alpha=0:V1.01:143GHz}{1.0148}
\setsymbol{HFI:colour:correction:alpha=0:V1.01:217GHz}{0.9909}
\setsymbol{HFI:colour:correction:alpha=0:V1.01:353GHz}{0.9888}
\setsymbol{HFI:colour:correction:alpha=0:V1.01:545GHz}{0.9878}
\setsymbol{HFI:colour:correction:alpha=0:V1.01:857GHz}{1.0014}

\providecommand{\sorthelp}[1]{}

%% file: PIP_118_Boulanger_authors_and_institutes.tex
\author{\small
Planck Collaboration: P.~A.~R.~Ade\inst{81}
\and
N.~Aghanim\inst{55}
\and
M.~Arnaud\inst{69}
\and
M.~Ashdown\inst{66, 5}
\and
J.~Aumont\inst{55}
\and
C.~Baccigalupi\inst{80}
\and
A.~J.~Banday\inst{89, 9}
\and
R.~B.~Barreiro\inst{61}
\and
N.~Bartolo\inst{28, 62}
\and
E.~Battaner\inst{91, 92}
% ...
}
\institute{\small
APC, AstroParticule et Cosmologie, Universit\'{e} Paris Diderot, CNRS/IN2P3, CEA/lrfu, Observatoire de Paris, Sorbonne Paris Cit\'{e}, 10, rue Alice Domon et L\'{e}onie Duquet, 75205 Paris Cedex 13, France\goodbreak
\and
Aalto University Mets\"{a}hovi Radio Observatory, P.O. Box 13000, FI-00076 AALTO, Finland\goodbreak
\and
African Institute for Mathematical Sciences, 6-8 Melrose Road, Muizenberg, Cape Town, South Africa\goodbreak
\and
Agenzia Spaziale Italiana Science Data Center, Via del Politecnico snc, 00133, Roma, Italy\goodbreak
\and
Astrophysics Group, Cavendish Laboratory, University of Cambridge, J J Thomson Avenue, Cambridge CB3 0HE, U.K.\goodbreak
\and
Astrophysics \& Cosmology Research Unit, School of Mathematics, Statistics \& Computer Science, University of KwaZulu-Natal, Westville Campus, Private Bag X54001, Durban 4000, South Africa\goodbreak
\and
Atacama Large Millimeter/submillimeter Array, ALMA Santiago Central Offices, Alonso de Cordova 3107, Vitacura, Casilla 763 0355, Santiago, Chile\goodbreak
\and
CITA, University of Toronto, 60 St. George St., Toronto, ON M5S 3H8, Canada\goodbreak
\and
CNRS, IRAP, 9 Av. colonel Roche, BP 44346, F-31028 Toulouse cedex 4, France\goodbreak
\and
California Institute of Technology, Pasadena, California, U.S.A.\goodbreak
\and
Centro de Estudios de F\'{i}sica del Cosmos de Arag\'{o}n (CEFCA), Plaza San Juan, 1, planta 2, E-44001, Teruel, Spain\goodbreak
\and
Computational Cosmology Center, Lawrence Berkeley National Laboratory, Berkeley, California, U.S.A.\goodbreak
\and
Consejo Superior de Investigaciones Cient\'{\i}ficas (CSIC), Madrid, Spain\goodbreak
\and
DSM/Irfu/SPP, CEA-Saclay, F-91191 Gif-sur-Yvette Cedex, France\goodbreak
\and
DTU Space, National Space Institute, Technical University of Denmark, Elektrovej 327, DK-2800 Kgs. Lyngby, Denmark\goodbreak
\and
D\'{e}partement de Physique Th\'{e}orique, Universit\'{e} de Gen\`{e}ve, 24, Quai E. Ansermet,1211 Gen\`{e}ve 4, Switzerland\goodbreak
\and
Departamento de Astrof\'{i}sica, Universidad de La Laguna (ULL), E-38206 La Laguna, Tenerife, Spain\goodbreak
\and
Departamento de F\'{\i}sica, Universidad de Oviedo, Avda. Calvo Sotelo s/n, Oviedo, Spain\goodbreak
\and
Department of Astrophysics/IMAPP, Radboud University Nijmegen, P.O. Box 9010, 6500 GL Nijmegen, The Netherlands\goodbreak
\and
Department of Physics \& Astronomy, University of British Columbia, 6224 Agricultural Road, Vancouver, British Columbia, Canada\goodbreak
\and
Department of Physics and Astronomy, Dana and David Dornsife College of Letter, Arts and Sciences, University of Southern California, Los Angeles, CA 90089, U.S.A.\goodbreak
\and
Department of Physics and Astronomy, University College London, London WC1E 6BT, U.K.\goodbreak
\and
Department of Physics, Florida State University, Keen Physics Building, 77 Chieftan Way, Tallahassee, Florida, U.S.A.\goodbreak
\and
Department of Physics, Gustaf H\"{a}llstr\"{o}min katu 2a, University of Helsinki, Helsinki, Finland\goodbreak
\and
Department of Physics, Princeton University, Princeton, New Jersey, U.S.A.\goodbreak
\and
Department of Physics, University of California, Santa Barbara, California, U.S.A.\goodbreak
\and
Department of Physics, University of Illinois at Urbana-Champaign, 1110 West Green Street, Urbana, Illinois, U.S.A.\goodbreak
\and
Dipartimento di Fisica e Astronomia G. Galilei, Universit\`{a} degli Studi di Padova, via Marzolo 8, 35131 Padova, Italy\goodbreak
\and
Dipartimento di Fisica e Scienze della Terra, Universit\`{a} di Ferrara, Via Saragat 1, 44122 Ferrara, Italy\goodbreak
\and
Dipartimento di Fisica, Universit\`{a} La Sapienza, P. le A. Moro 2, Roma, Italy\goodbreak
\and
Dipartimento di Fisica, Universit\`{a} degli Studi di Milano, Via Celoria, 16, Milano, Italy\goodbreak
\and
Dipartimento di Matematica, Universit\`{a} di Roma Tor Vergata, Via della Ricerca Scientifica, 1, Roma, Italy\goodbreak
\and
Discovery Center, Niels Bohr Institute, Blegdamsvej 17, Copenhagen, Denmark\goodbreak
\and
European Southern Observatory, ESO Vitacura, Alonso de Cordova 3107, Vitacura, Casilla 19001, Santiago, Chile\goodbreak
\and
European Space Agency, ESAC, Planck Science Office, Camino bajo del Castillo, s/n, Urbanizaci\'{o}n Villafranca del Castillo, Villanueva de la Ca\~{n}ada, Madrid, Spain\goodbreak
\and
European Space Agency, ESTEC, Keplerlaan 1, 2201 AZ Noordwijk, The Netherlands\goodbreak
\and
Facolt\`{a} di Ingegneria, Universit\`{a} degli Studi e-Campus, Via Isimbardi 10, Novedrate (CO), 22060, Italy\goodbreak
\and
Finnish Centre for Astronomy with ESO (FINCA), University of Turku, V\"{a}is\"{a}l\"{a}ntie 20, FIN-21500, Piikki\"{o}, Finland\goodbreak
\and
Gran Sasso Science Institute, INFN, viale F. Crispi 7, 67100 L'Aquila, Italy\goodbreak
\and
HGSFP and University of Heidelberg, Theoretical Physics Department, Philosophenweg 16, 69120, Heidelberg, Germany\goodbreak
\and
Helsinki Institute of Physics, Gustaf H\"{a}llstr\"{o}min katu 2, University of Helsinki, Helsinki, Finland\goodbreak
\and
INAF - Osservatorio Astrofisico di Catania, Via S. Sofia 78, Catania, Italy\goodbreak
\and
INAF - Osservatorio Astronomico di Padova, Vicolo dell'Osservatorio 5, Padova, Italy\goodbreak
\and
INAF - Osservatorio Astronomico di Roma, via di Frascati 33, Monte Porzio Catone, Italy\goodbreak
\and
INAF - Osservatorio Astronomico di Trieste, Via G.B. Tiepolo 11, Trieste, Italy\goodbreak
\and
INAF/IASF Bologna, Via Gobetti 101, Bologna, Italy\goodbreak
\and
INAF/IASF Milano, Via E. Bassini 15, Milano, Italy\goodbreak
\and
INFN, Sezione di Bologna, Via Irnerio 46, I-40126, Bologna, Italy\goodbreak
\and
INFN, Sezione di Roma 1, Universit\`{a} di Roma Sapienza, Piazzale Aldo Moro 2, 00185, Roma, Italy\goodbreak
\and
INFN, Sezione di Roma 2, Universit\`{a} di Roma Tor Vergata, Via della Ricerca Scientifica, 1, Roma, Italy\goodbreak
\and
IPAG: Institut de Plan\'{e}tologie et d'Astrophysique de Grenoble, Universit\'{e} Grenoble Alpes, IPAG, F-38000 Grenoble, France, CNRS, IPAG, F-38000 Grenoble, France\goodbreak
\and
Imperial College London, Astrophysics group, Blackett Laboratory, Prince Consort Road, London, SW7 2AZ, U.K.\goodbreak
\and
Infrared Processing and Analysis Center, California Institute of Technology, Pasadena, CA 91125, U.S.A.\goodbreak
\and
Institut Universitaire de France, 103, bd Saint-Michel, 75005, Paris, France\goodbreak
\and
Institut d'Astrophysique Spatiale, CNRS (UMR8617) Universit\'{e} Paris-Sud 11, B\^{a}timent 121, Orsay, France\goodbreak
\and
Institut d'Astrophysique de Paris, CNRS (UMR7095), 98 bis Boulevard Arago, F-75014, Paris, France\goodbreak
\and
Institute of Astronomy, University of Cambridge, Madingley Road, Cambridge CB3 0HA, U.K.\goodbreak
\and
Institute of Theoretical Astrophysics, University of Oslo, Blindern, Oslo, Norway\goodbreak
\and
Instituto Nacional de Astrof\'{\i}sica, \'{O}ptica y Electr\'{o}nica (INAOE), Apartado Postal 51 y 216, 72000 Puebla, M\'{e}xico\goodbreak
\and
Instituto de Astrof\'{\i}sica de Canarias, C/V\'{\i}a L\'{a}ctea s/n, La Laguna, Tenerife, Spain\goodbreak
\and
Instituto de F\'{\i}sica de Cantabria (CSIC-Universidad de Cantabria), Avda. de los Castros s/n, Santander, Spain\goodbreak
\and
Istituto Nazionale di Fisica Nucleare, Sezione di Padova, via Marzolo 8, I-35131 Padova, Italy\goodbreak
\and
Jet Propulsion Laboratory, California Institute of Technology, 4800 Oak Grove Drive, Pasadena, California, U.S.A.\goodbreak
\and
Jodrell Bank Centre for Astrophysics, Alan Turing Building, School of Physics and Astronomy, The University of Manchester, Oxford Road, Manchester, M13 9PL, U.K.\goodbreak
\and
Kavli Institute for Cosmological Physics, University of Chicago, Chicago, IL 60637, USA\goodbreak
\and
Kavli Institute for Cosmology Cambridge, Madingley Road, Cambridge, CB3 0HA, U.K.\goodbreak
\and
LAL, Universit\'{e} Paris-Sud, CNRS/IN2P3, Orsay, France\goodbreak
\and
LERMA, CNRS, Observatoire de Paris, 61 Avenue de l'Observatoire, Paris, France\goodbreak
\and
Laboratoire AIM, IRFU/Service d'Astrophysique - CEA/DSM - CNRS - Universit\'{e} Paris Diderot, B\^{a}t. 709, CEA-Saclay, F-91191 Gif-sur-Yvette Cedex, France\goodbreak
\and
Laboratoire Traitement et Communication de l'Information, CNRS (UMR 5141) and T\'{e}l\'{e}com ParisTech, 46 rue Barrault F-75634 Paris Cedex 13, France\goodbreak
\and
Laboratoire de Physique Subatomique et Cosmologie, Universit\'{e} Grenoble-Alpes, CNRS/IN2P3, 53, rue des Martyrs, 38026 Grenoble Cedex, France\goodbreak
\and
Laboratoire de Physique Th\'{e}orique, Universit\'{e} Paris-Sud 11 \& CNRS, B\^{a}timent 210, 91405 Orsay, France\goodbreak
\and
Lawrence Berkeley National Laboratory, Berkeley, California, U.S.A.\goodbreak
\and
Lebedev Physical Institute of the Russian Academy of Sciences, Astro Space Centre, 84/32 Profsoyuznaya st., Moscow, GSP-7, 117997, Russia\goodbreak
\and
Max-Planck-Institut f\"{u}r Astrophysik, Karl-Schwarzschild-Str. 1, 85741 Garching, Germany\goodbreak
\and
National University of Ireland, Department of Experimental Physics, Maynooth, Co. Kildare, Ireland\goodbreak
\and
Nicolaus Copernicus Astronomical Center, Bartycka 18, 00-716 Warsaw, Poland\goodbreak
\and
Niels Bohr Institute, Blegdamsvej 17, Copenhagen, Denmark\goodbreak
\and
Optical Science Laboratory, University College London, Gower Street, London, U.K.\goodbreak
\and
SISSA, Astrophysics Sector, via Bonomea 265, 34136, Trieste, Italy\goodbreak
\and
School of Physics and Astronomy, Cardiff University, Queens Buildings, The Parade, Cardiff, CF24 3AA, U.K.\goodbreak
\and
Simon Fraser University, Department of Physics, 8888 University Drive, Burnaby BC, Canada\goodbreak
\and
Sorbonne Universit\'{e}-UPMC, UMR7095, Institut d'Astrophysique de Paris, 98 bis Boulevard Arago, F-75014, Paris, France\goodbreak
\and
Space Research Institute (IKI), Russian Academy of Sciences, Profsoyuznaya Str, 84/32, Moscow, 117997, Russia\goodbreak
\and
Space Sciences Laboratory, University of California, Berkeley, California, U.S.A.\goodbreak
\and
Special Astrophysical Observatory, Russian Academy of Sciences, Nizhnij Arkhyz, Zelenchukskiy region, Karachai-Cherkessian Republic, 369167, Russia\goodbreak
\and
Sub-Department of Astrophysics, University of Oxford, Keble Road, Oxford OX1 3RH, U.K.\goodbreak
\and
UPMC Univ Paris 06, UMR7095, 98 bis Boulevard Arago, F-75014, Paris, France\goodbreak
\and
Universit\'{e} de Toulouse, UPS-OMP, IRAP, F-31028 Toulouse cedex 4, France\goodbreak
\and
Universities Space Research Association, Stratospheric Observatory for Infrared Astronomy, MS 232-11, Moffett Field, CA 94035, U.S.A.\goodbreak
\and
University of Granada, Departamento de F\'{\i}sica Te\'{o}rica y del Cosmos, Facultad de Ciencias, Granada, Spain\goodbreak
\and
University of Granada, Instituto Carlos I de F\'{\i}sica Te\'{o}rica y Computacional, Granada, Spain\goodbreak
% ...
}

%% file: DustEB.bbl
\def\eprinttmppp@#1arXiv:@{#1}
\providecommand{\arxivlink[1]}{\href{http://arxiv.org/abs/#1}{arXiv:#1}}
\def\eprinttmp@#1arXiv:#2 [#3]#4@{\ifthenelse{\equal{#3}{x}}{\ifthenelse{
\equal{#1}{}}{\arxivlink{\eprinttmppp@#2@}}{\arxivlink{#1}}}{\arxivlink{#2}
  [#3]}}
\providecommand{\eprintlink}[1]{\eprinttmp@#1arXiv: [x]@}
\providecommand{\eprint}[1]{\eprintlink{#1}}
\providecommand{\adsurl}[1]{\href{#1}{ADS}}